\documentclass[aps,twocolumn,preprintnumbers,nofootinbib,superscriptaddress]{revtex4-1}
\usepackage{eurosym}
\usepackage{amsmath}
\usepackage{graphicx}
\usepackage{amsfonts}
\usepackage{array}
\usepackage{amsthm}
\usepackage{bm}
\usepackage{palatino}
\usepackage{mathpazo}
\usepackage{supertabular}
\usepackage{subfig}
\usepackage[breaklinks]{hyperref}
\usepackage{graphicx}
\usepackage{caption}
\usepackage[font=small,labelfont=bf,justification=raggedright]{caption}
\usepackage{color}
\usepackage{booktabs}
\usepackage{multirow}
\begin{document}

\title{Imprints of Dark Matter on Black Hole Shadows using Spherical Accretions}
\author{Saurabh}
\email{sbhkmr1999@gmail.com}
\affiliation{Department of Physics, \\Dyal Singh College, University of Delhi-110003}
\author{Kimet Jusufi}
\email{kimet.jusufi@unite.edu.mk (corresponding author)}
\affiliation{Physics Department, State University of Tetovo, Ilinden Street nn, 1200,
Tetovo, North Macedonia}
\begin{abstract}
 We study the possibility of identifying dark matter in the galactic center from the  physical properties of the electromagnetic radiation emitted from an optically-thin disk region around a static and spherically symmetric black hole. In particular, we consider two specific models for the optical-thin disk region: a gas at rest and a gas in a radial free fall.  Due to the effect of dark matter on the spacetime geometry, we find that the dark matter can increase or decrease the intensity of the electromagnetic flux radiation depending on the dark matter model.  To this end,  we analyze two simple dark matter models having different mass functions $\mathcal{M}(r)$, with a matter mass $M$, thickness $\Delta r_s$ along with a dark matter core radius surrounding the black hole. In addition to that, we explore the scenario of a perfect fluid dark matter surrounding the black hole.  We show that in order to have significant effect of dark matter on the intensity of the electromagnetic flux radiation, a high energy density of dark matter near the black hole is needed.  We also find that the surrounding dark matter distribution plays a key role on the shadow radius and the intensity of the electromagnetic flux radiation, respectively. Finally we have used the relation between the shadow radius and the quasinormal modes (QNMs) to compute the real part of QNM frequencies.
\end{abstract}

\maketitle

\section{Introduction}
Many modern astrophysical observatories point towards the fact that giant elliptical and spiral galaxies contain supermassive black holes (SMBHs) at their galactic centers. For example, observations suggest that at the center of our Milky Way galaxy there is a supermassive black hole with four million solar masses. These supermasssive black holes are characterized by huge masses and spin parameter (or angular momenta).
	According to Einstein's theory of relativity,  black holes generically contain a spacetime singularity at their center and an event horizon where the gravity is so strong that nothing even the electromagnetic radiation or light—can escape from it.  Due to the exterior spacetime geometry, black hole (BH) can capture light received from nearby stars or accretion disks into bound orbits. In other words, every black hole is characterized by the photon sphere or a  collection of light  rays orbiting the BH \cite{Cunha:2020azh}. In particular, the orbit of light is said to be unstable if the photon can fall into the BH or escape to infinity. The most compelling evidence of the existence of black holes is the the first shadow images of the SMBH at the center of M87 galaxy  Event Horizon Telescope (EHT) collaboration has detected [2,3] and the detection of gravity waves by LIGO [4]. It is interesting that, by monitoring of motion of stars in the Galactic Center and by collecting precise measurements of the orbital motion of such stars we can probe on the nature of black holes and the surrounding spacetime \cite{Do:2019txf,Naoz:2019sjx,Do:2008uf}.\\
	
	It's quite amazing that we can use these precise observations to constrain different physically viable BH solutions and therefore we can test General Relativity and alternative theories of gravity, say by observing small deviations from the Kerr solution. Toward this goal, we can explore the distortion in the shadow images which encodes valuable information about the the black hole mass/spin, and also the spacetime geometry around a given black hole solution. Furthermore, the shadow images can be used to test the existence of other exotic objects such as wormholes and naked singularities. 	In literature, one can find many detailed studies concerning the black hole shadows, including static and spherically symmetric solutions, rotating solutions, and solutions in different theories of gravity [8-58]. For further studies concerning the effect of black hole geometry on the electromagnetic radiation see Refs. \cite{f1,f2,f3,f4}.
	
	 From the astrophysical point of view, it is well known that dark matter which is assumed to be some form of elementary particle plays a key role in many astrophysical processes, yet is one of the greatest unsolved mysteries.  In recent papers, the effect of dark matter on black hole shadow has been investigated [66-68]. In this paper, we aim to  study the possibility of identifying dark matter in the galactic center based on the  physical properties of the electromagnetic radiation emitted from a optical-thin disk region around the static and spherically symmetric black hole. Furthermore, we shall consider a radiating optically-thin disk of gas at rest and a radiating thin disk of gas in a free fall. 
	
	The structure of our paper is laid out as follows: In Section II, firstly we review the dark matter model proposed in Ref. \cite{Konoplya:2019sns} having dark matter with positive and negative energy density around the black hole.  In Section III and Section IV, we study the shadow images and intensity of the radiation produced by a spherically thin medium described by a gas at rest and infalling gas model, respectively. In Section V, we study a perfect fluid dark matter model surrounding a black hole and the shadow images/intensity of the radiation produced by a gas at rest and an infalling gas model, respectively. In Section VI, we explore the connection between the shadow radius and the quasinormal modes (QNMs). Finally, in Section VII, we comment on our results. 
\section{Black hole surrounded by dark matter}
\subsection{Model I}
Let us start by considering a toy model of a Schwarzschild black hole surrounded by dark matter proposed recently in \cite{Konoplya:2019sns}.  One can then assume a piecewise function to impose three domains \cite{Konoplya:2019sns}:
\begin{eqnarray} \label{massfunc}
\mathcal{M}(r)=\begin{cases}
m, & r<r_{s};\\
m+ M \mathcal{G}(r), & r_{s}\leq r\leq r_{s} +\Delta r_{s};\\
m+ M, & r>r_{s}+\Delta r_{s}
\end{cases}
\end{eqnarray}
where
\begin{equation} \label{e2}
\mathcal{G}(r)=\left(3-2\frac{r-r_{s}}{\Delta r_{s}}\right)\left(\frac{r-r_{s}}{\Delta r_{s}}\right)^{2}.
\end{equation}
The expression for $\mathcal{G}(r)$ is chosen so that $\mathcal{M}(r)$ and $\mathcal{M}'(r)$ are continuous (see Fig. 1). 
The Schwarzschild metric surrounded by a spherical shell of dark matter described by Eq. \eqref{e2} can be written as follows
\begin{equation} 
ds^2 = -f(r)dt^2 +f(r)^{-1}dr^2 +r^2\left(d\theta^2+\sin^2 \theta d\phi^2\right)
\end{equation}
in which the metric function $f(r)$ reads
\begin{equation} \label{e4}
f(r)=1-\frac{2\mathcal{M}(r)}{r}.
\end{equation}

\begin{figure}
   	\includegraphics[width=7.5 cm]{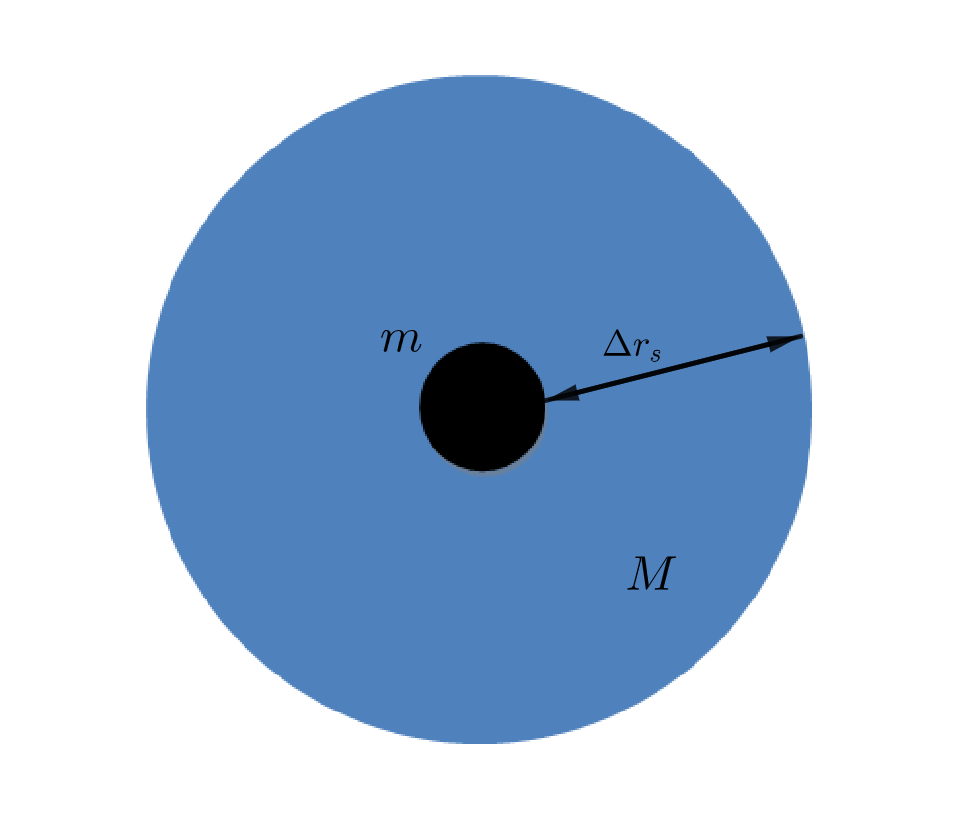}
    \caption{A black hole with mass $m$ surrounded by 
dark matter shell with thickness $\Delta r_s$ and mass $M$. Note that $r_s=r_h$. }
\end{figure}

In the present work, we shall consider the nontrivial case where the dark matter is concentrated near the black hole such that  $r_{s}= r_{h}$, as shown in Fig. 1. The mass function then is given by $m+ M \mathcal{G}(r)$ along with $r_{s}\leq r\leq r_{s} +\Delta r_{s}.$ From the field equations, we can compute the energy-momentum components 
\begin{eqnarray}
{T^{t}}_{t}={T^{r}}_{r}=-\frac{\mathcal{M}'(r)}{4 \pi r^2},
\end{eqnarray}
and
\begin{eqnarray}
{T^{\theta}}_{\theta}={T^{\phi}}_{\phi}=-\frac{\mathcal{M}''(r)}{8 \pi r}.
\end{eqnarray}

Note that we are going to consider two cases of the surrounding dark matter near the black hole; namely, a dark matter with positive energy density, i.e. when $M>0$ (normal matter), and a dark matter with negative energy density i.e. $M<0$. Previously, the shadow images of metric (3) and rotating counterpart of metric (3) have been studied in Refs. [66,67].

\begin{figure*}[t]
  \centering
  \includegraphics[scale=0.3]{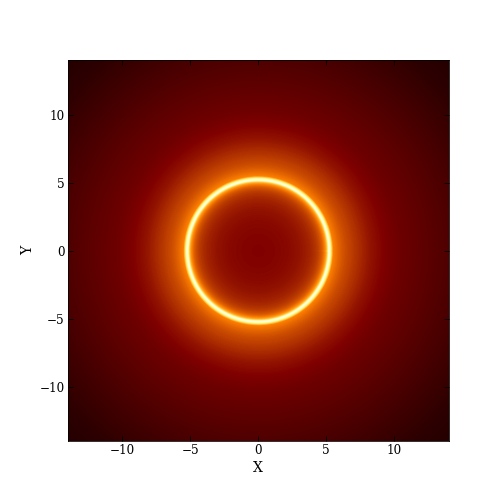}
  \includegraphics[scale=0.3]{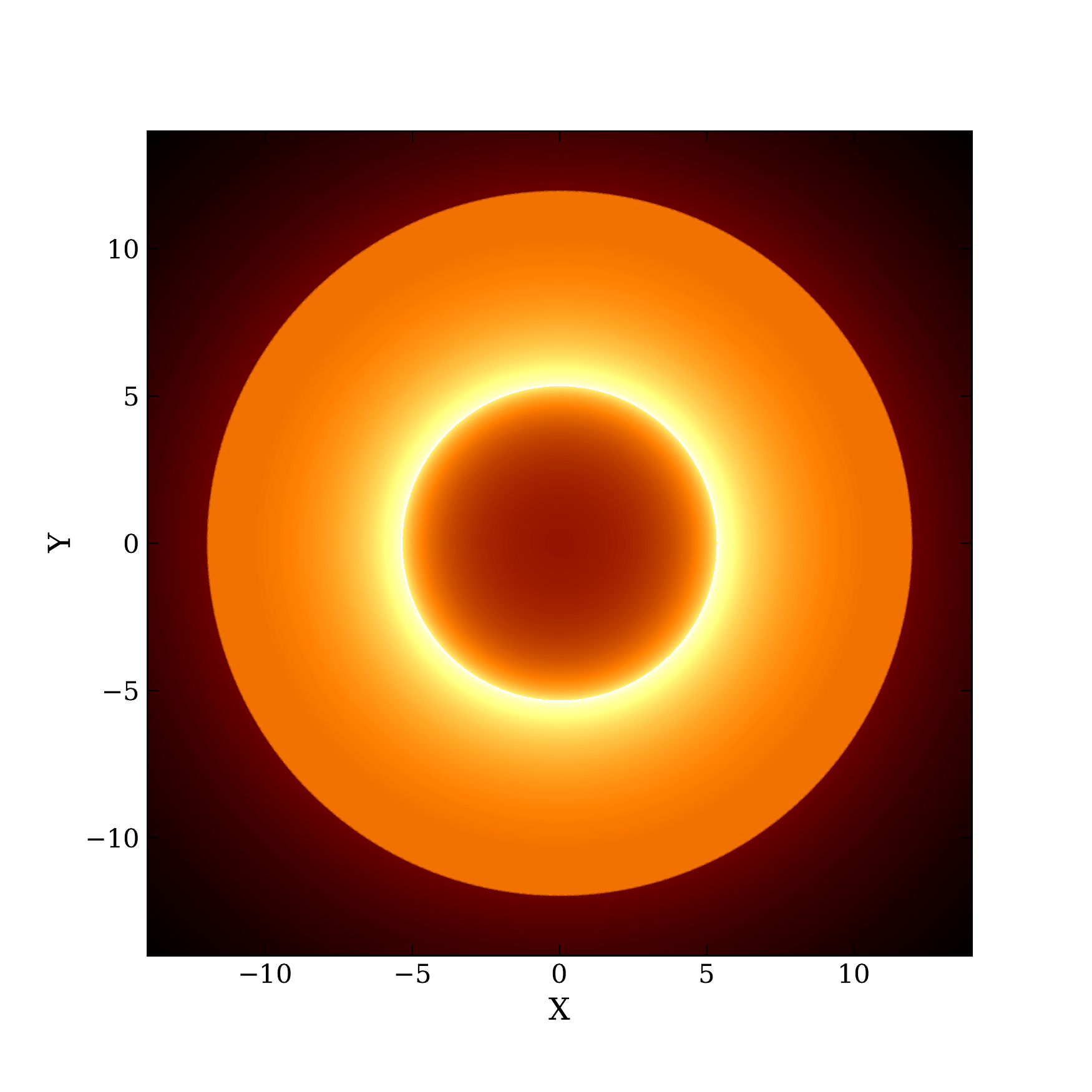}
  \includegraphics[scale=0.3]{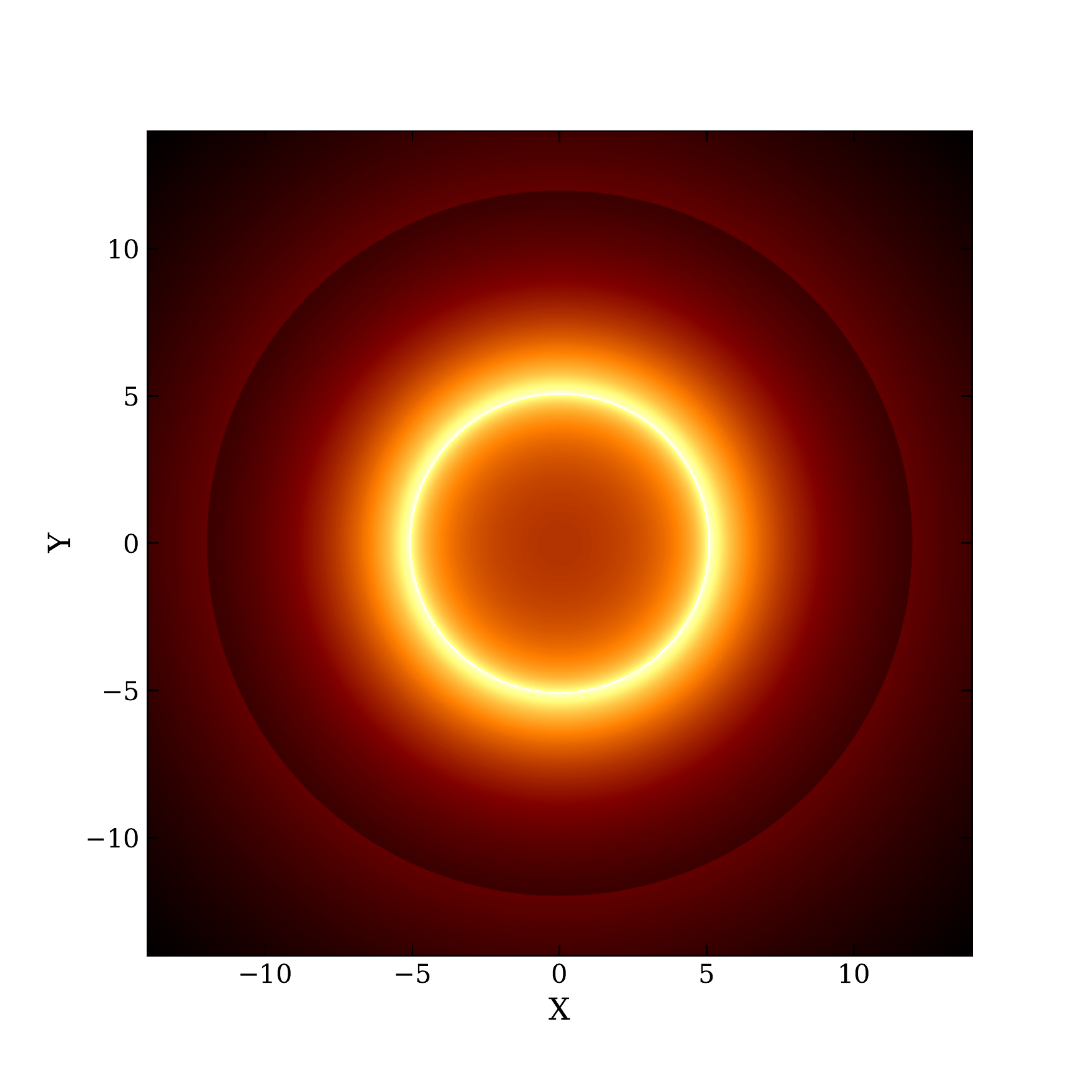}
  \includegraphics[scale=0.3]{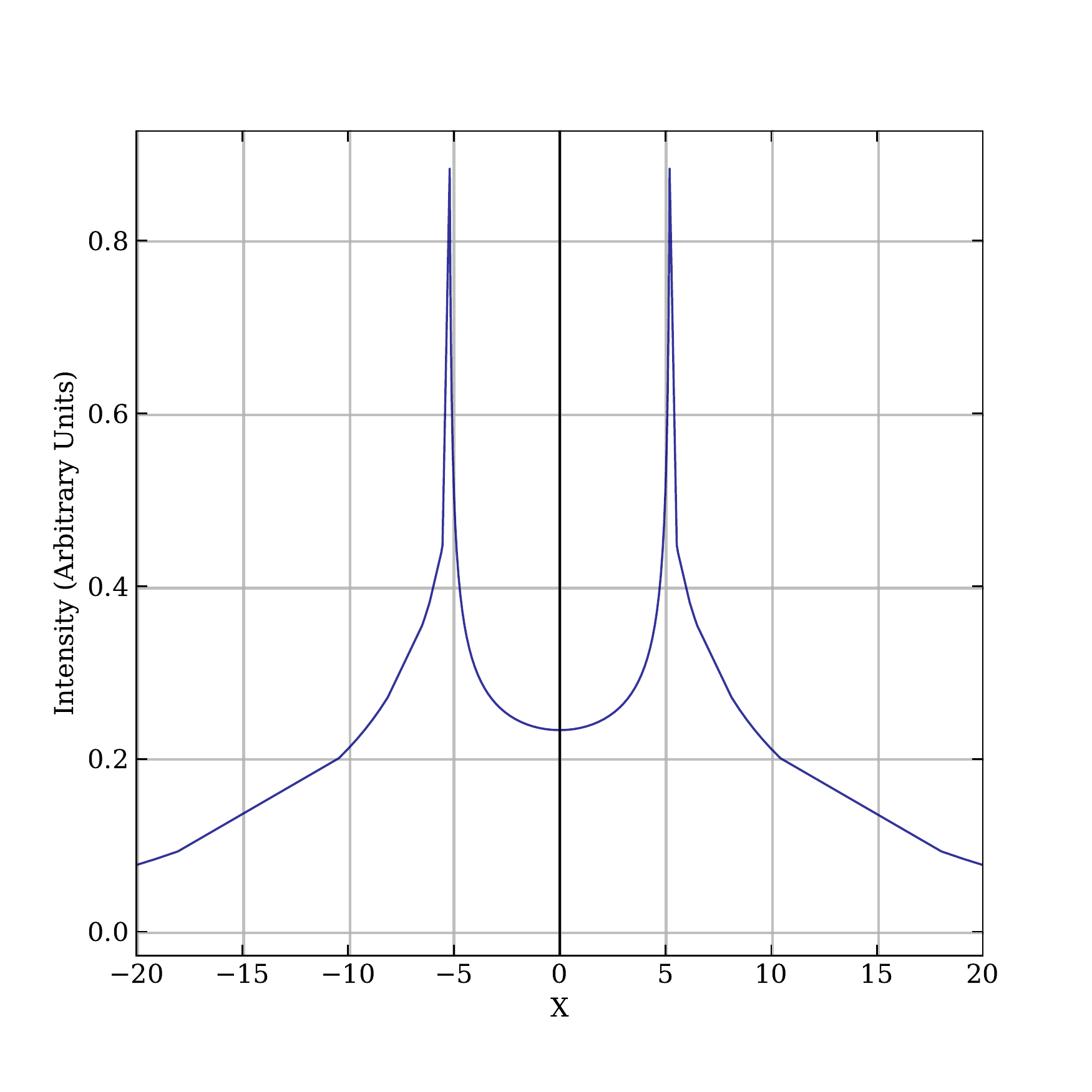}
  \includegraphics[scale=0.3]{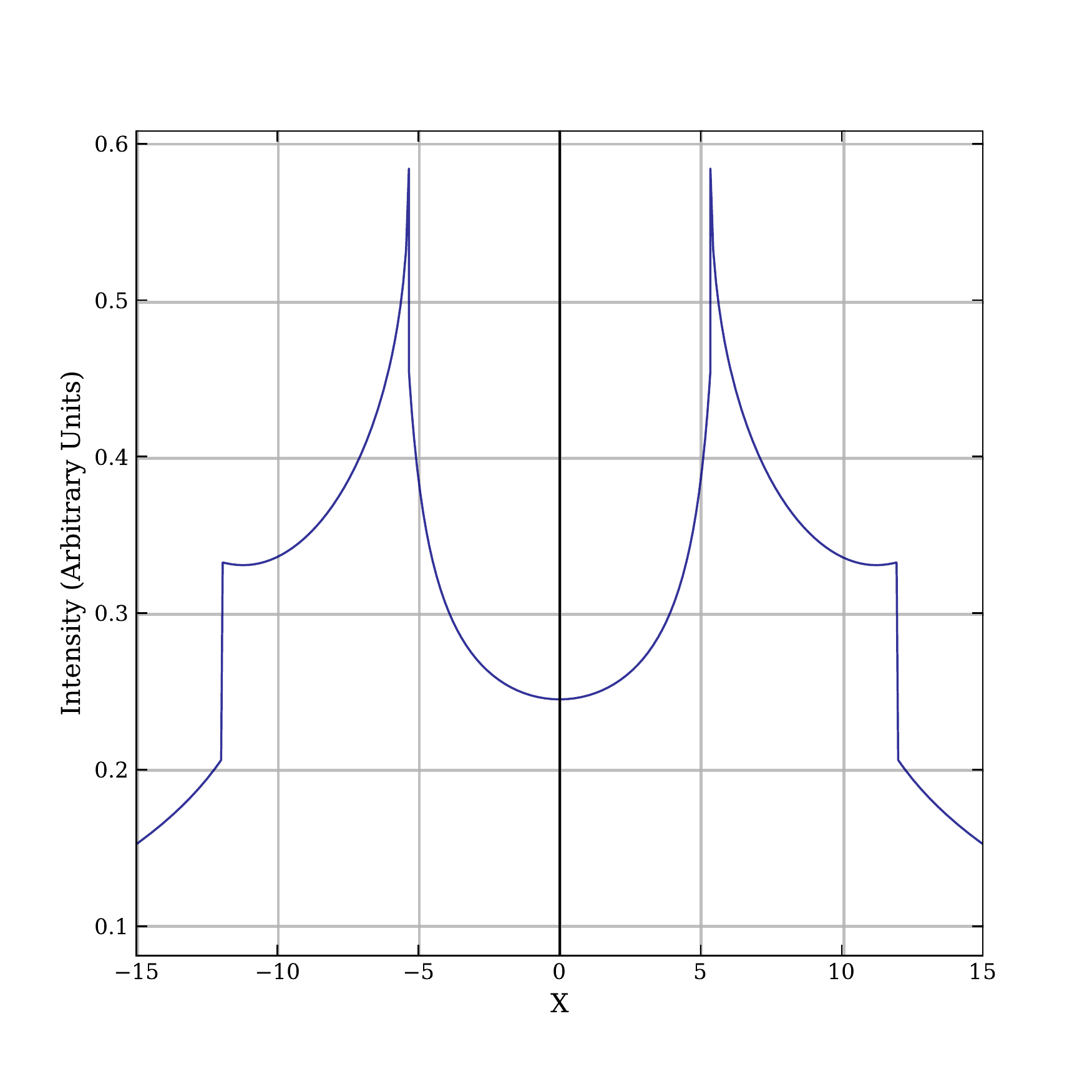}
  \includegraphics[scale=0.3]{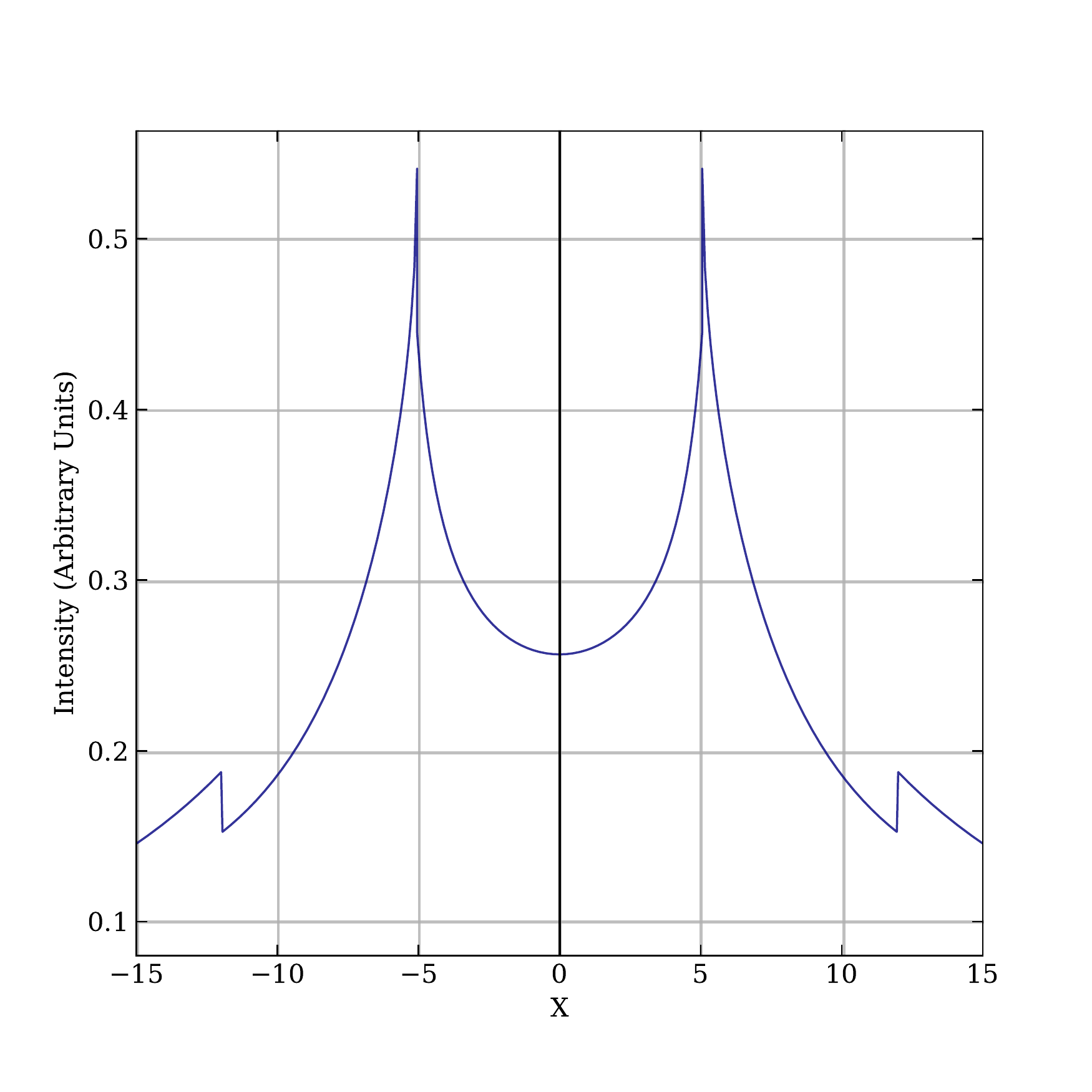}
  \caption{From left to right: Image of shadows and the corresponding intensities using the rest gas model seen by a distant observer for the Schwarzschild black hole, black hole surrounded by dark matter having positive energy density $M>0$ and negative energy density $M<0$, respectively. We have chosen $m=1$, $M=1m$ along with $r_s=2 m$ and $\Delta r_s=10 m$. }
\end{figure*}
\begin{figure*}[t]
  {\includegraphics[scale=0.3]{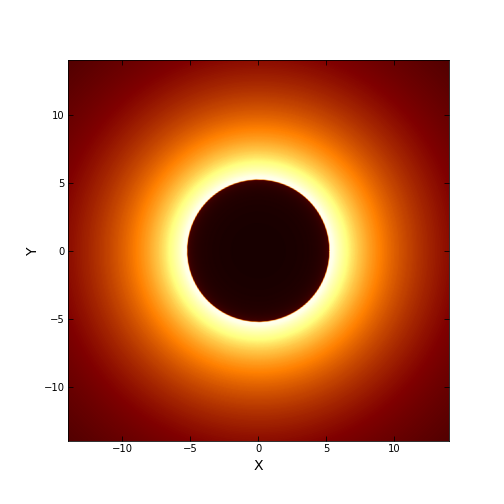}}
  {\includegraphics[scale=0.3]{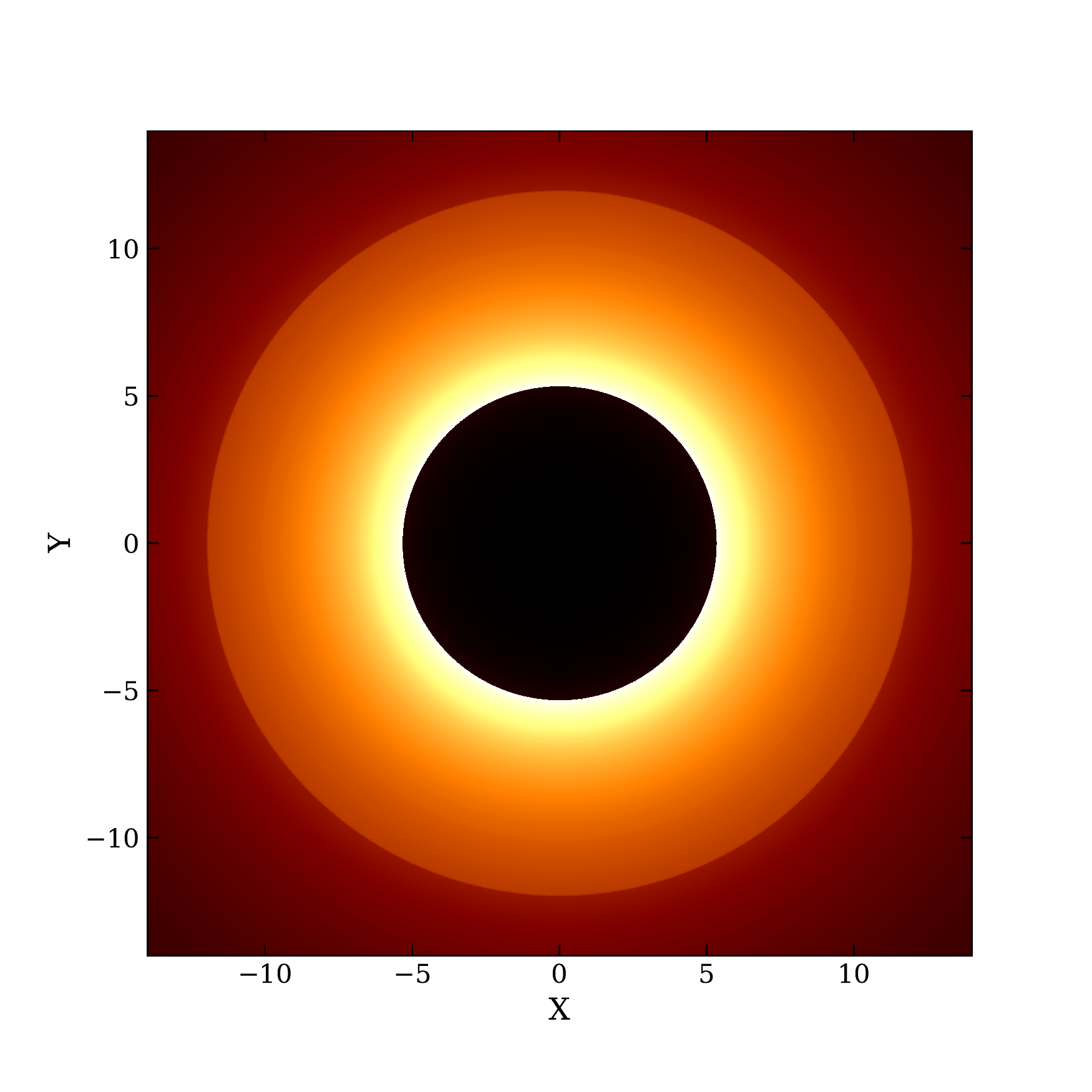}}
  {\includegraphics[scale=0.3]{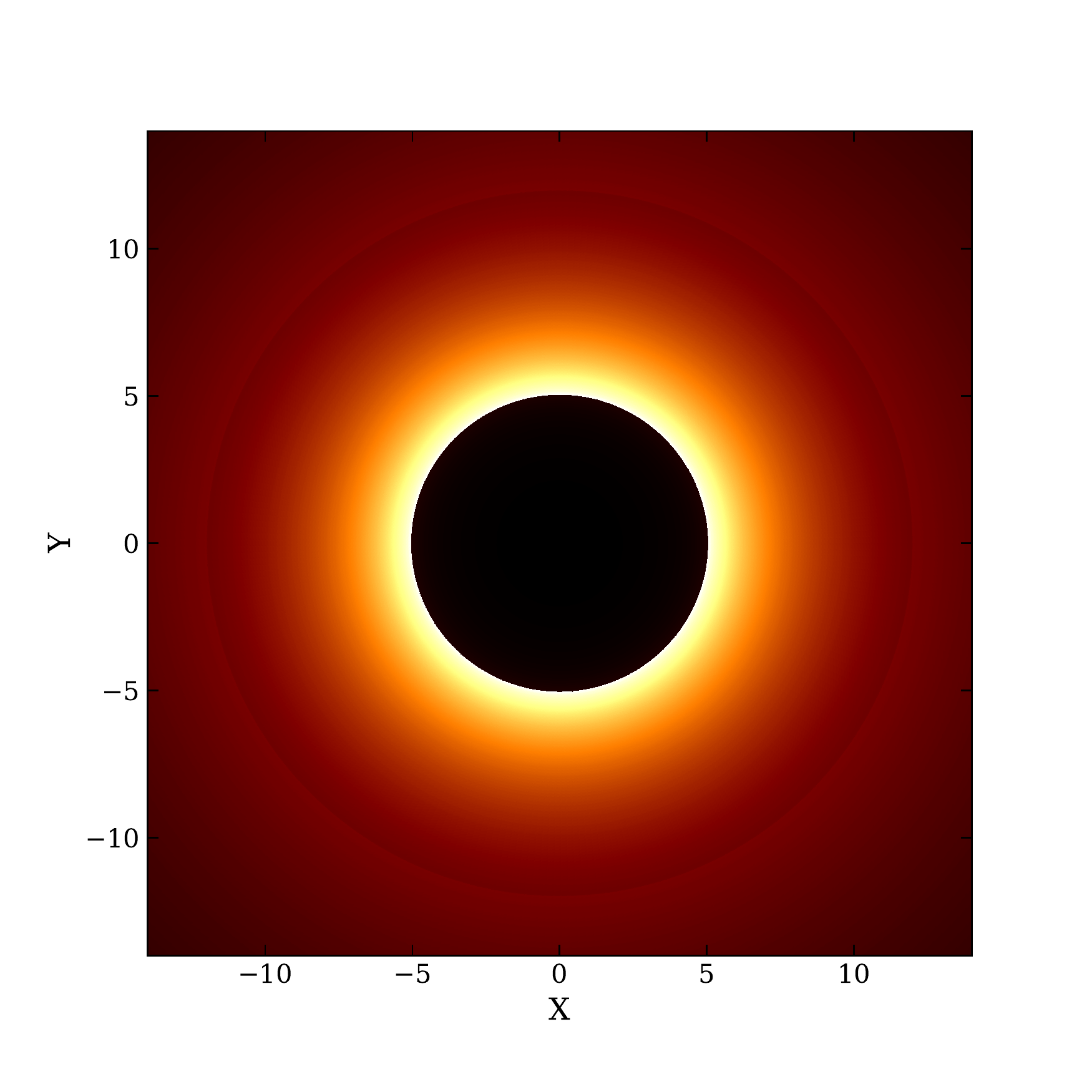}}
  {\includegraphics[scale=0.3]{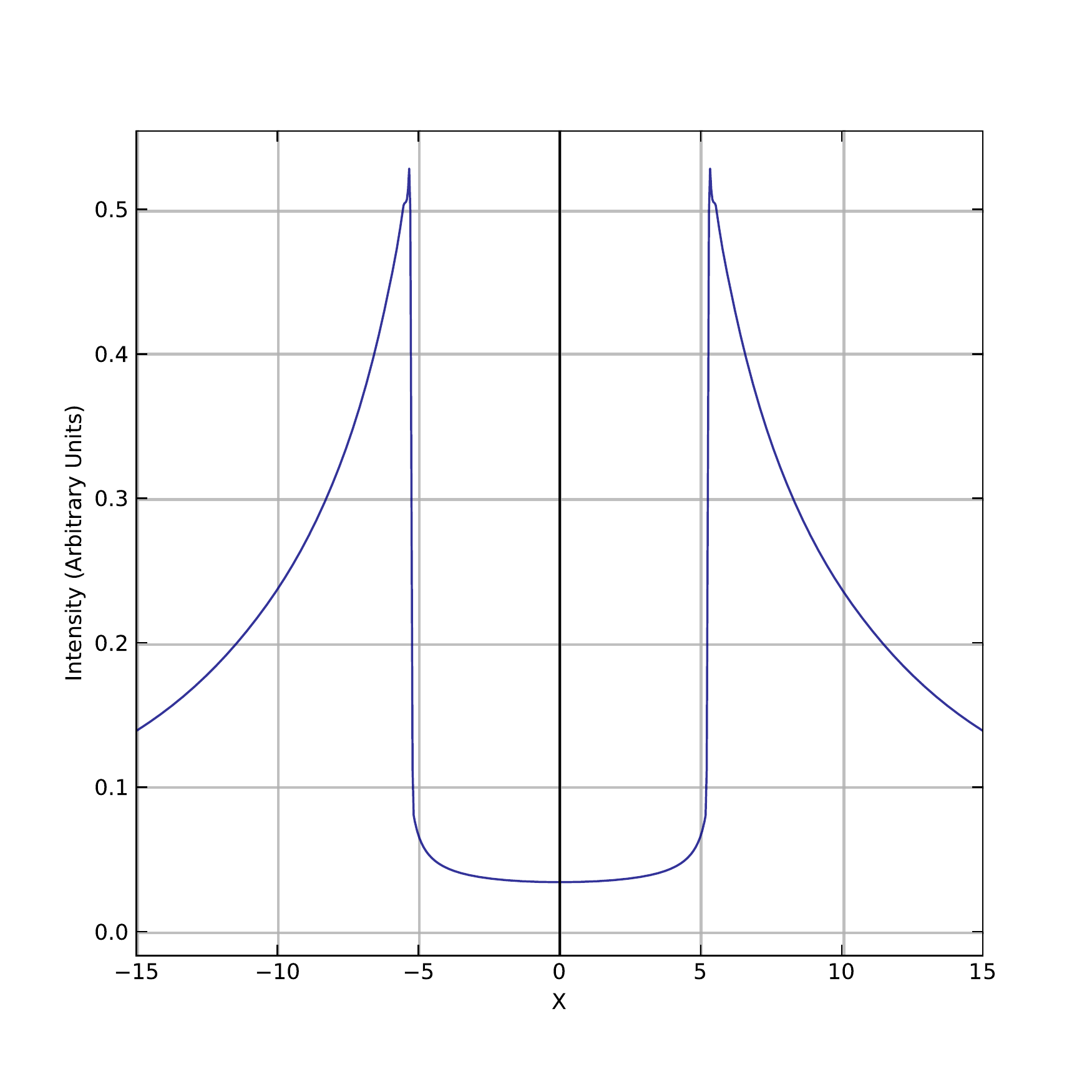}}
  {\includegraphics[scale=0.3]{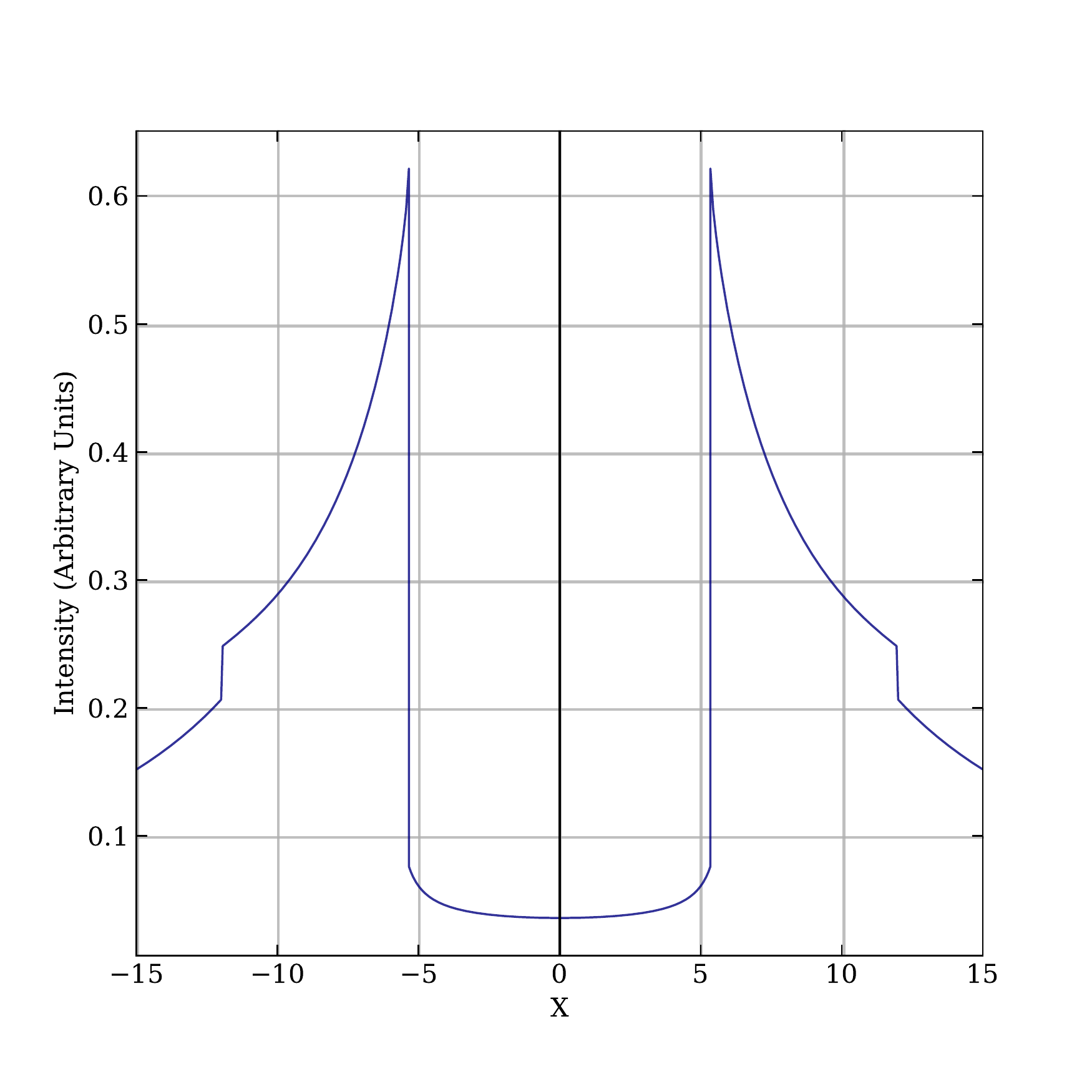}}
  {\includegraphics[scale=0.3]{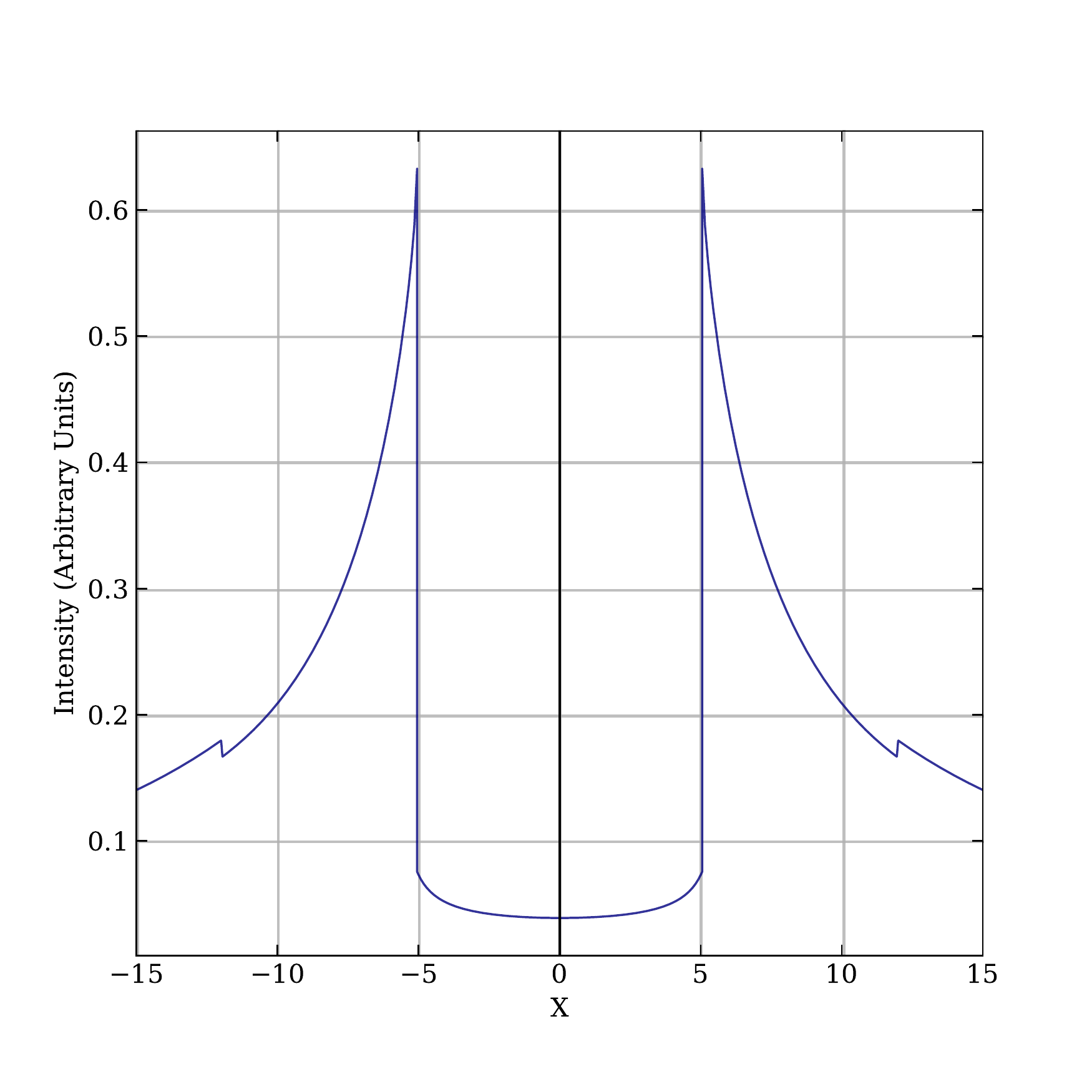}}
\caption{From the left to the right: Image of shadows and the corresponding intensities using the infalling gas model seen by a distant observer for the Schwarzschild black hole, black hole surrounded by dark matter having positive energy density $M>0$ and negative energy density $M<0$, respectively. We have chosen $m=1$, $M=1m$ along with $r_s=2 m$ and $\Delta r_s=10 m$. }
\end{figure*}


\subsection{Model II: Black hole in perfect fluid dark matter }
In order to have a more realistic model for dark matter we need to take into account not only the energy density but also the relativistic pressure of dark matter.  To do so, let us consider a $3+1$ gravity theory in the presence of a perfectly fluid dark matter (PFDM) with the action [70,71,72] (see also Refs. \cite{Xu:2018wow})
\begin{eqnarray}
S=\int d^4x \sqrt{-g} \left( \frac{ R}{16 \pi} +\mathcal{L}_{DM}\right).
\end{eqnarray}
From the action, we get the Einstein field equations as
\begin{eqnarray}
R_{\mu\nu}+\frac{1}{2}g_{\mu \nu}R=8 \pi T_{\mu \nu}^{DM}
\end{eqnarray}
where $T_{\mu \nu}^{DM}$ corresponds to the energy-momentum
tensor of the perfect fluid dark matter with components $ (T^{\mu}_{\nu})^{DM}={\rm diag}(-\rho, P_r, P, P)$. Let us now investigate a static and a spherically symmetric solutions, for that we need to employ the following metric
\begin{equation}
ds^2 = -e^\nu dt^2 + e^{\lambda} dr^2 + r^2 (d\theta^2 + \sin^2\theta d\phi^2),
\end{equation}
adding the ansatz, $\nu=-\lambda$, from  the Einsteins field equations we find
\begin{eqnarray}
&& e^{-\lambda} \left( \frac{1}{r^2} - \frac{\lambda^\prime}{r} \right) -\frac{1}{r^2} =\kappa^2 T^t_{\ t} \,, \label{eq:R-1}\\
&& e^{-\lambda} \left( \frac{1}{r^2} + \frac{\nu^\prime}{r} \right) -\frac{1}{r^2} = \kappa^2 T^r_{\ r} \,,  \label{eq:R-2}\\
&& \frac{e^{-\lambda}}{2} \left( \nu{''} +  \frac{\nu'^2}{2}   + \frac{\nu{'} -\lambda{'}}{r} - \frac{\nu^\prime \lambda^\prime}{2} \right)  = \kappa^2 T^\theta_{\ \theta}  \,, \label{eq:R-3}\\
&& \frac{e^{-\lambda}}{2} \left( \nu{''} +  \frac{\nu'^2}{2}   + \frac{\nu{'} -\lambda{'}}{r} - \frac{\nu^\prime \lambda^\prime}{2}  \right)  = \kappa^2 T^\phi_{\ \phi} \,.\label{eq:R-4}
\end{eqnarray}
and a {\it prime} denotes  the differentiation with respect to $r$. If we set $T^\theta_{\ \theta}=T^\phi_{\ \phi}= T^t_{\ t}(1-\epsilon)$ with $\epsilon$ being a constant yielding \cite{Xu:2017bpz})
\begin{eqnarray}
\frac{P}{\rho}=\epsilon-1.
\end{eqnarray}

It is interesting to note that we can find solutions that are satisfied with the condition $\lambda=-\nu$, {\it i.e.}  $g_{tt} = -g_{rr}^{-1}$,  in the limit $T^t_{\ t} \to  T^r_{\ r} $.  By using the equation of state for the PFDM to find the solution satisfied with the condition $\lambda=-\nu$,  using  $\nu= \ln (1-U)$  with $U=U(r)$ along with the field equations it can be found \cite{Xu:2018wow}
 \begin{eqnarray}
&& r^2 U'' + 2 \epsilon\, r U' +2 (\epsilon-1) U=0.
\end{eqnarray}

We can consider two special cases of the above equation.

\subsubsection{Case $\epsilon=3/2$}
In this particular case it was found the following solution \cite{Li,Xu:2018wow}
\begin{eqnarray}
 U=\frac{1}{r}  \left[r_{Sh} -  a \ln \left( \frac{r}{|a|} \right) \right],  \ \  {\rm for}\  \epsilon = \frac{3}{2}  \,,
  \end{eqnarray}
where $r_{Sh}=2m$. Then the corresponding black hole spacetime in perfect fluid dark matter is given by \cite{Li,K1}
\begin{equation}
ds^2=-f(r)dt^2+\frac{dr^2}{f(r)}+r^2\left(d\theta^2+\sin^2\theta \,d\phi^2\right),
\end{equation}
with
\begin{equation}
f(r)=1-\frac{2m}{r}+\frac{a}{r}\ln\left(\frac{r}{|a|}\right),
\end{equation}
where $m$ is the black hole mass and $a$ is a parameter describing the intensity of the dark matter.  In the present work, we are going to assume $a>0$. Hence we can write the following approximation
\begin{equation}
e^{\ln\left(\frac{r}{|a|}\right)^{\frac{a}{r}}}= \left( \frac{r}{|a|} \right)^{\frac{a}{r}}\simeq 1+ \frac{a}{r}\ln\left(\frac{r}{|a|}\right),
\end{equation}
if we identify $a \to 2 v_0^2 r_0$, under this scaling we obtain in the large limit $r$
\begin{eqnarray}
f(r) \simeq 1+2 v^2(r)\ln\left(\frac{r}{2 v_0^2 r_0}\right)
\end{eqnarray}
and the tangential velocity is a function of $r$, $v^2(r)=v_0^2 r_0/r$. 
In order to see the physical interpretation of $v_0$ let us consider the isothermal dark matter density profile given by the following relation
\begin{eqnarray}
\rho(r)=\frac{v_0^2}{4 \pi r^2},
\end{eqnarray}
along with the mass profile of the dark matter galactic halo given by
\begin{equation}
M_{DM}\left( r\right) =4\pi \int_{0}^{r}\rho
\left( r'\right) r^{'2}dr'=v_0^2 r,
\end{equation}
can explain the flat curve of motions of stars in the galaxy.  To see this, let us use the last equation to find the tangential velocity $v_{tg}^{2}\left( r\right) =M_{DM}(r)/r=v_0^2$ which describes a region with the nearly constant tangential speed of the stars.  To do so, we can work with a static and spherically symmetric spacetime ansatz with pure dark matter in Schwarzschild coordinates  the tangential velocity, one can calculate the radial function $F(r)$ by the following equation \cite{Xu:2018wow}
\begin{align}
      v_{tg}^{2}\left( r\right) =\frac{r}{\sqrt{F(r)}}\frac{d \sqrt{F(r)}}{dr}=r \frac{d\ln(\sqrt{F(r)})}{dr}.
\end{align}

Assuming a spherically symmetric solution, and solving the last equation we find
\begin{equation}
F(r)=\left(\frac{r}{r_0}\right)^{2 v_0^2 } 
\end{equation}
where $r_0$ is some constant. From the pure dark matter space-time it was argued that one can obtain the space-time metric of a black hole surrounded by dark matter halo solving the Einstein field equations (8) and using ${T^{\nu}_{~\mu}}^{DM}=diag[-\rho,p_{r},p,p]$ for non-zero energy-momentum corresponding to the pure dark matter space-time metric. One way to include the black hole in our metric is by treating the dark matter as part of the general energy-momentum tensor ${T_{\mu \nu}}^{DM}$.  The space-time metric is chosen as (9)  yielding a coefficient functions as follows  (see for details \cite{Xu:2018wow})
\begin{equation}
f(r)=\left(\frac{r}{r_0}\right)^{2 v_0^2 }- \dfrac{2 m}{r}.
\end{equation}

Thus, the above space-time has a black hole in the dark matter halo. Finally, the metric coefficient 
$f(r)$ reduces to $f(r)\simeq 1+2 v_0^2 \ln(r/r_0)$ which explains the flat curve of motions of stars in the galaxy having $v_0 \simeq 10^{-3}$. In other words, the last metric function describes the spacetime in the regime where the dark matter effect is not very strong and it is a special case of the metric function (18) having constant velocity $v_0$.  Thus,  the metric function (18) describes a strong regime of dark matter effect near the black hole. 

\subsubsection{Case $\epsilon \not= \frac{3}{2} $}
The solution of  equation (15) in this case reads \cite{Li}
 \begin{eqnarray}
 U=\frac{r_{Sh}}{r}  - \frac{r^{2(1-\epsilon)}}{r_\epsilon},  \ \  {\rm for}\  \epsilon \not= \frac{3}{2}  \,,
  \end{eqnarray}
along with the corresponding metric given by
\begin{eqnarray}
ds^2 = - f(r)dt^2  +  \frac{dr^2}{f(r)}
+r^2(d\theta^2 +\sin^2\theta d\phi^2) \,,  \label{eq:metric-1}
\end{eqnarray}
with
\begin{eqnarray}
f(r)=\left[ 1-\frac{2m}{r}  + \frac{r^{2(1-\epsilon)}}{r_\epsilon}\right].
\end{eqnarray}

In this paper, we are interested in the metric that can describe the motions of stars in the galaxy. Therefore let us consider the scaling 
\begin{equation}
    \alpha \to 2 (1-\epsilon),\,\,\,\,r_\epsilon^{-1}=\gamma/r_0^\alpha,
\end{equation}
then the metric function (28) gives
  \begin{equation}
 f=1- \frac{2m}{r} +\gamma \Bigg( \frac{r}{r_0} \Bigg)^\alpha.
\end{equation}
This metric describes the extreme case of the dark-matter galaxy with the SMBH at its center and show the corresponding metric function $-g_{tt}$ as a function of $r$. Furthermore we can consider a Taylor series expansion around $\alpha$ far away from the black hole  yielding 
  \begin{equation}
 f(r) \simeq 1+\gamma \left(1+\alpha \ln(\frac{r}{r_0})\right)
\end{equation}
where it was introduced $\gamma \alpha \simeq 2 v_0^2$. In what follows we shall consider the shadow images using the dark matter models that we elaborated in this section.


\section{Photon sphere and the shadow radius}
Here, we are interested in investigating the shadow of black hole solution surrounded by matter. To do so, we start from Hamilton-Jacobi method for null geodesics in the black hole spacetime written as \cite{Perlick:2015vta}
\begin{equation}
\frac{\partial S}{\partial \sigma}+H=0,
\end{equation}
in which $S $ is the Jacobi action and  $\sigma $  is some affine parameter along
the geodesics.  If we consider a photon along null geodesics in our spherically symmetrical spacetime surrounded by matter, one can show that the Hamiltonian can be written as
\begin{equation}
\frac{1}{2}\left[-\frac{p_{t}^{2}}{f(r)}+f(r)p_{r}^{2}+\frac{p_{\phi}^{2}}{r^{2}}\right] =0.
\label{EqNHa}
\end{equation}
Due to the spacetime symmetries related to the coordinates $t$ and $\phi$, there are two constants of motion defined $p_{t}=-E$ and $p_{\phi}=L$, where $E$ and $L$ are the energy and the angular momentum of the photon, respectively. Next, the circular and unstable orbits are related to the the maximum value of effective potential in terms of the following conditions
\begin{equation}
V_{\rm eff}(r) \big \vert_{r=r_{ph}}=0,  \qquad \frac{\partial V_{\rm eff}(r)}{\partial r}%
\Big\vert_{r=r_{ph}}=0,  
\end{equation}
Without going into details here, one can now show the following equation of motion 
\begin{equation}
\frac{dr}{d\phi}=\pm r\sqrt{f(r)\left[\frac{r^{2}f(R)}{R^{2}f(r)} -1\right] }. 
\end{equation}
Let us consider a light ray sent from a static observer located at a position $r_{0} $ and transmitted with an angle $\vartheta$ with respect to the radial direction. We, therefore, have \cite{Perlick:2015vta}
\begin{equation}
\cot \vartheta =\frac{\sqrt{g_{rr}}}{g_{\phi\phi}}\frac{dr}{d\phi}\Big\vert%
_{r=r_{0}}.  \label{Eqangle}
\end{equation}
Finally,  the relation for shadow radius of the black hole as observed by a static observer at the position $r_0$  can be shown as 
\begin{equation}
R_{s}=r_{0}\sin \vartheta =R\sqrt{\frac{f(r_{0})}{f(R)}}\Bigg\vert_{R=r_{ph}}.
\end{equation}
where $r_{ph}$ represents the photon sphere radius and  $ r_{0} $ is the position of the observer located at a far distance from the black hole. We note here that the when the spacetime is assumed to be asymptotically flat, for the location of the observer in the region far away from the black hole it follows that $f(r_0)=1$. In our case, the metric function (4) and (18) are asymptotically flat, but the metric function (30) in general is not asymptotically flat. In our model, the dark matter can be located inside the photon sphere. therefore, in general, dark matter modifies the photon radius and one should be careful in choosing the correct photon sphere, i.e.  the one with the
smallest impact parameter. Concerning model I, it was shown that there are two solutions for the photon sphere with opposite sign and this was already analyzed by \cite{Konoplya:2019sns}. On the other hand, for the perfect fluid dark matter case one can obtain the solution for the photon sphere only numerically. 
\newpage

\begin{figure*}[t]
\centering
  \includegraphics[scale=0.36]{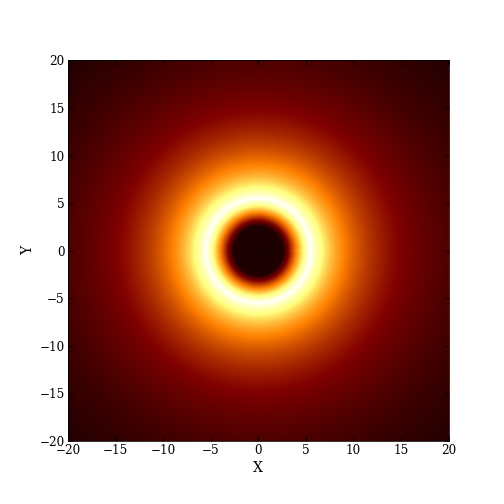}
  \includegraphics[scale=0.36]{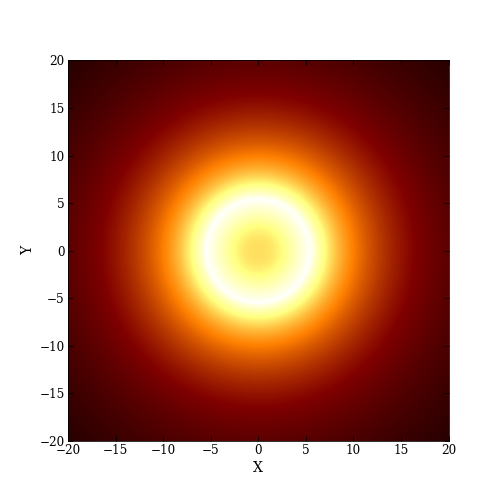}
  \includegraphics[scale=0.36]{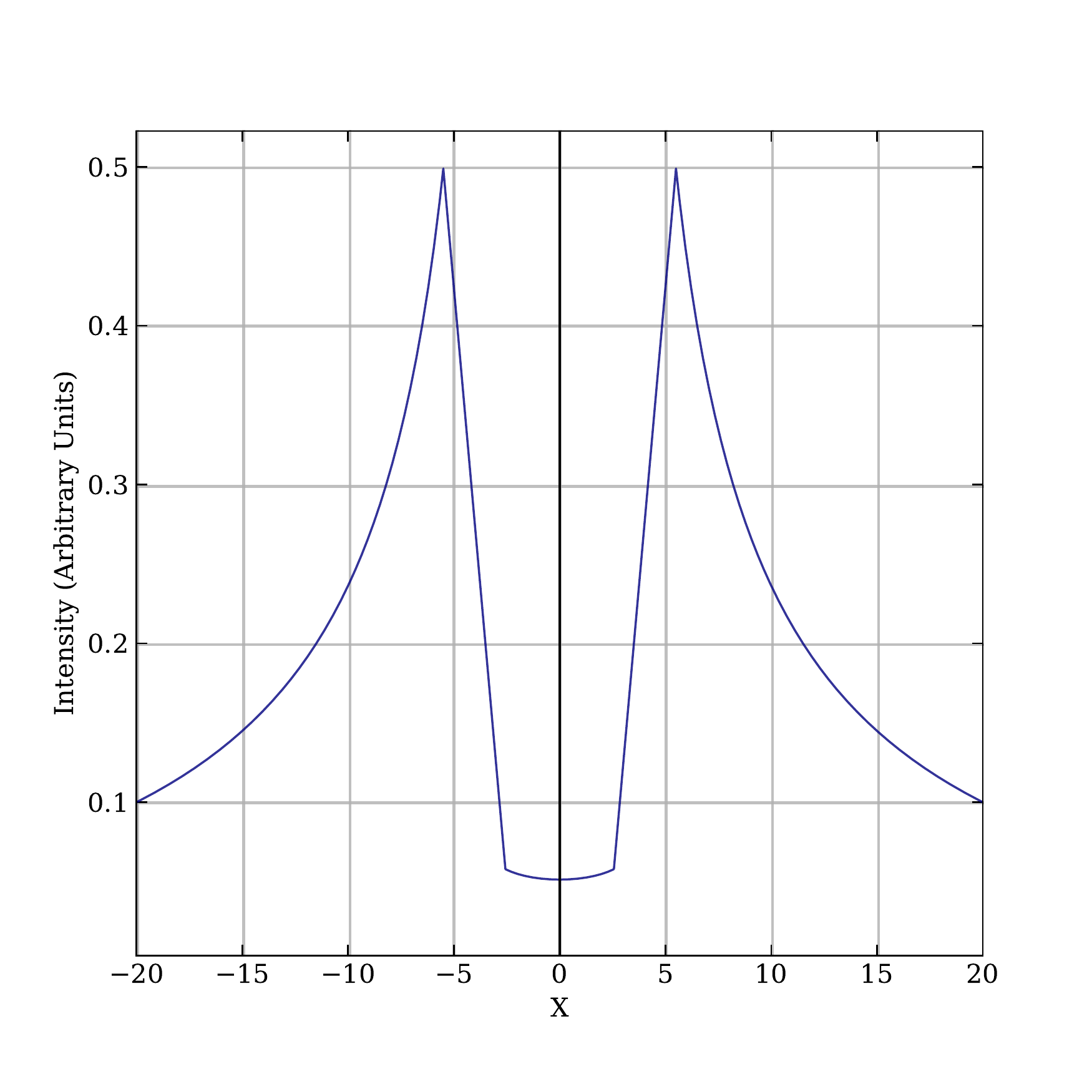}
   \includegraphics[scale=0.36]{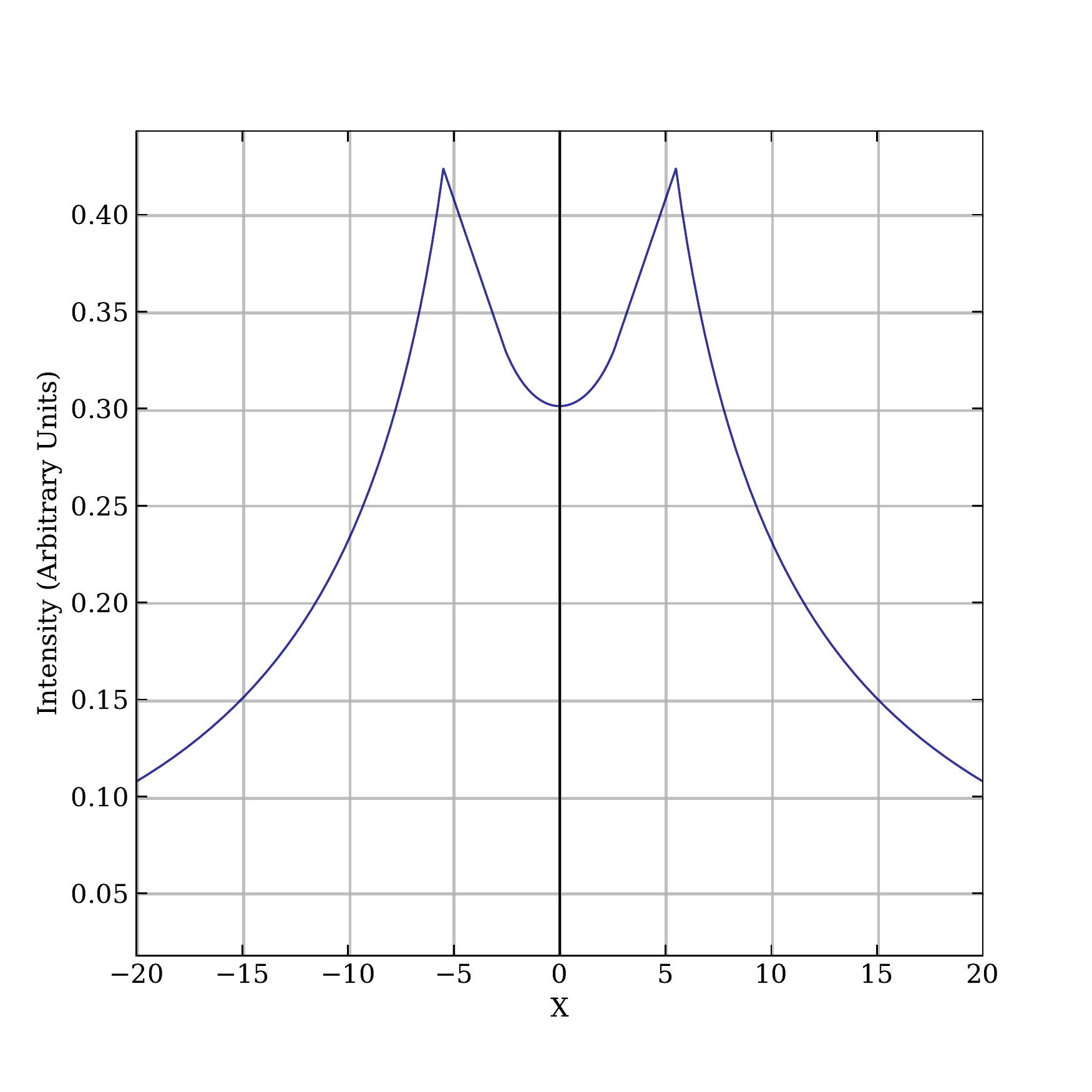}
\caption{Left panel: Images of shadows and the corresponding intensities using the infalling gas model in the spacetime of a black hole in perfect fluid dark matter with the metric function (18). Right panel: Images of shadows and the corresponding intensities using the static gas model in perfect fluid dark matter with the metric function (18).  We have chosen $a=0.1$ in both plots, respectively.}
\end{figure*}

\begin{figure*}[t]
  \centering
  {\includegraphics[scale=0.36]{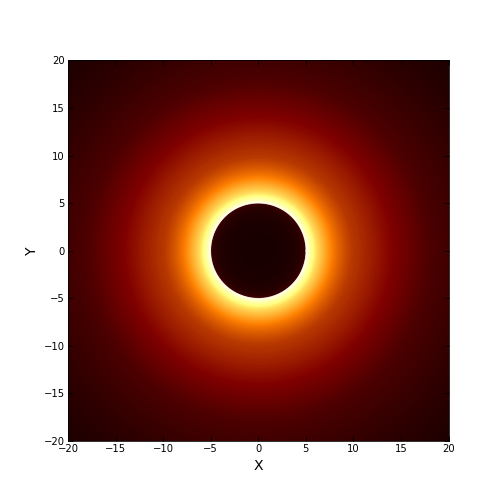}}
  {\includegraphics[scale=0.36]{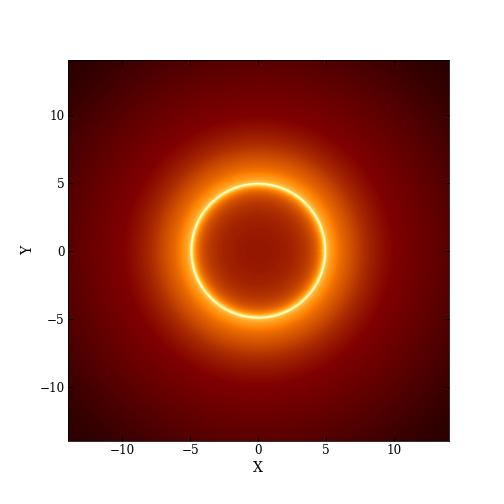}}
  {\includegraphics[scale=0.36]{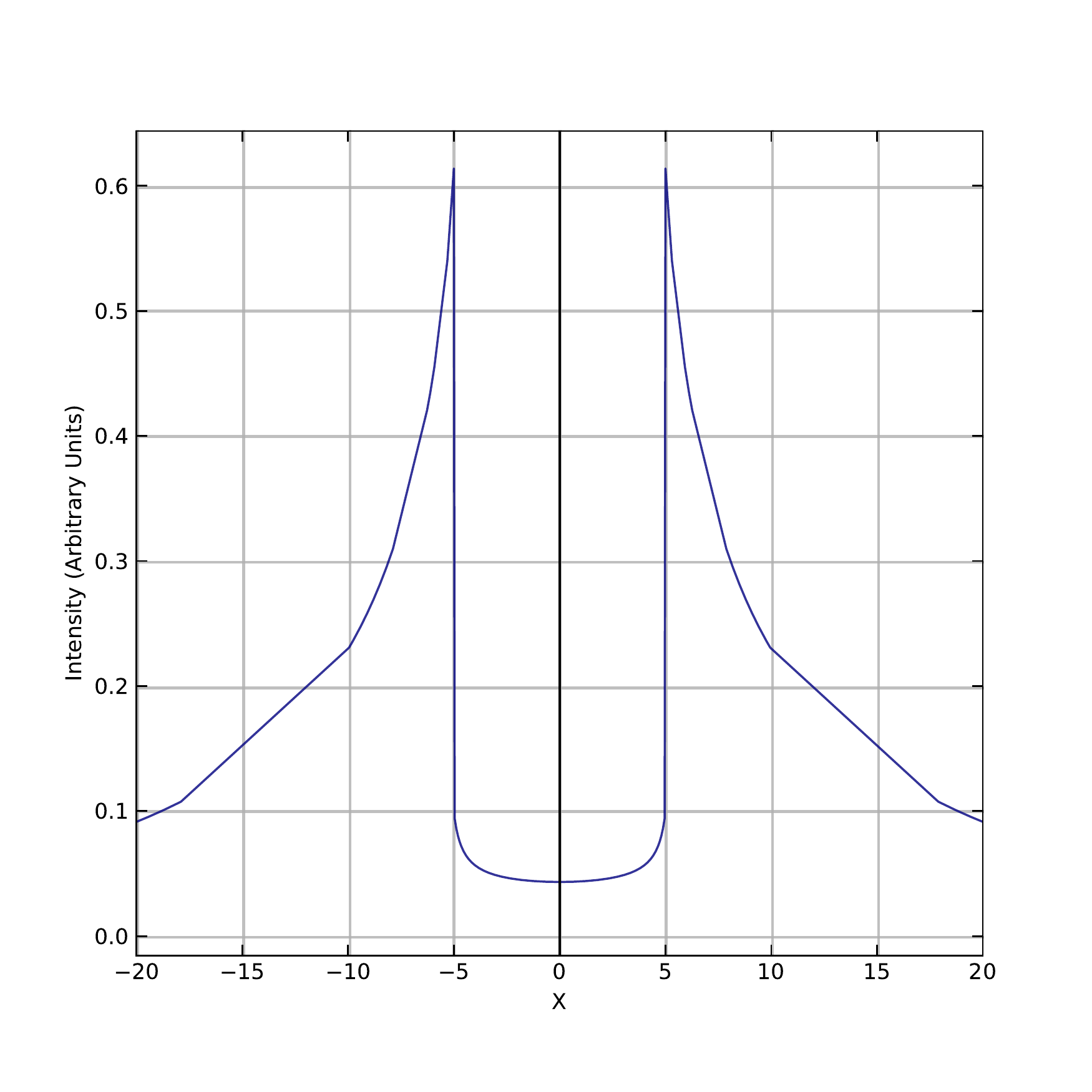}}  
  {\includegraphics[scale=0.36]{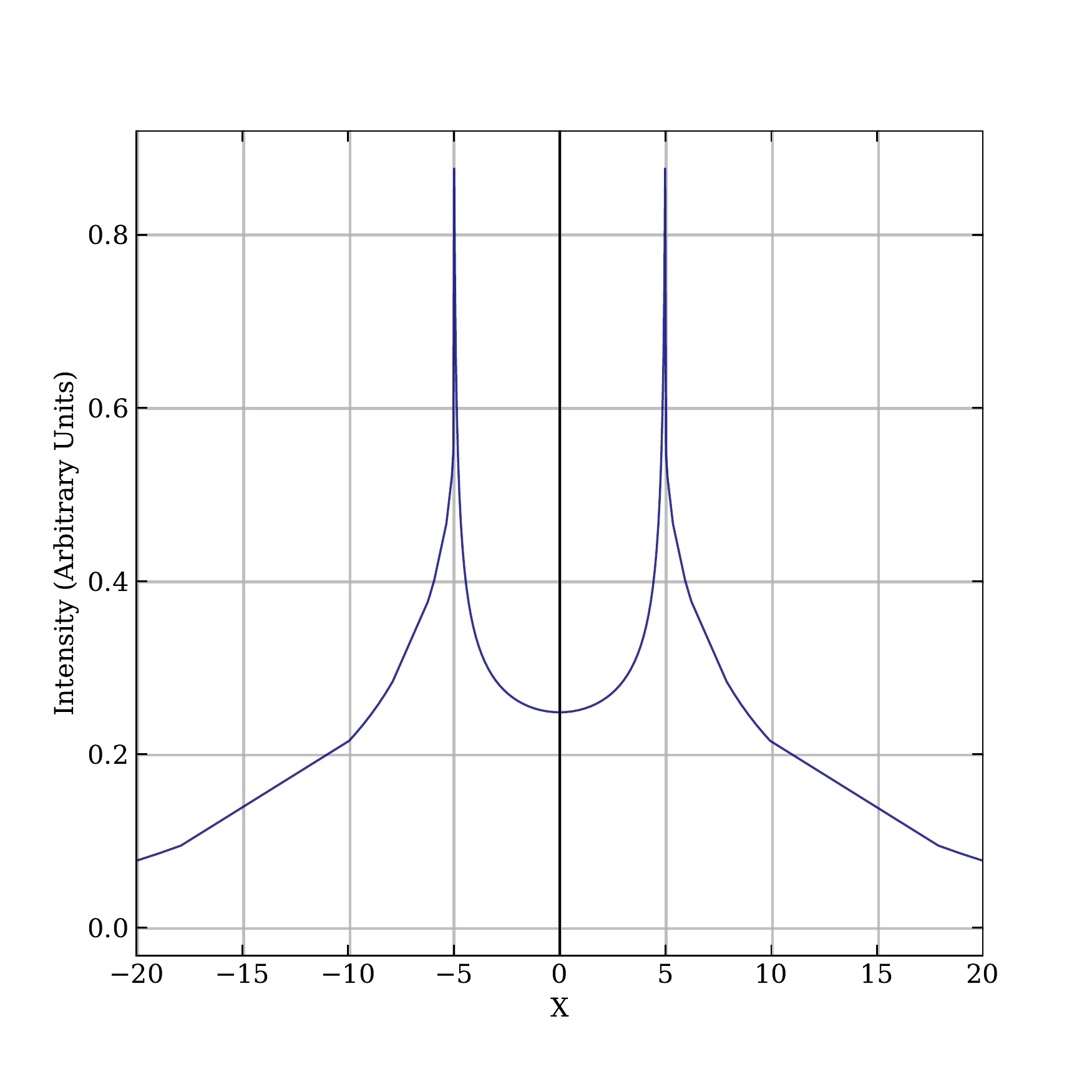}}
\caption{Left panel: Image of shadow and the corresponding intensity using the infalling gas model in the spacetime of a black hole in perfect fluid dark matter with the metric function (30). Right panel: Image of shadow and the corresponding intensity using the rest gas model in the spacetime of a black hole in perfect fluid dark matter with the metric function (30). We have chosen $\alpha = 0.1, \gamma = 0.1, r_0 = 10^6$ in both plots.}
\end{figure*}

\section{Optically thin radiating gas at rest surrounding the black hole}
Let us start by considering a general relativistic analysis for a thin radiating gas at rest in the spacetime of black hole surrounded by dark matter. To do so, we shall consider the following \cite{Narayan}: (i) The proper length corresponding
to the coordinate interval $dr$ is $\sqrt{1/f(r)} dr$. (ii) Every unit
of energy emitted in the local frame at $r$ corresponds to
$\sqrt{f(r)}$ units of energy at infinity. (iii) A time interval dt
in the local frame corresponds to an interval $dt/\sqrt{f(r)}$
at infinity. It follows we define the power per
unit volume $P_s$, and the emission coefficient per unit solid angle
$j_s$ in terms of the following relation \cite{Narayan}:
\begin{eqnarray}
P_s(r)=\frac{1}{|g_{tt}| \sqrt{g_{rr}} 4 \pi r^4}=4 \pi j_s(r),
\end{eqnarray}
where in our case we have $g_{tt}=f(r)$ along with  $g_{rr}=1/f(r)$. The net emitted luminosity in this model, measured at infinity, reads
\begin{eqnarray}
L_{emit}=\int_{r}^{\infty}|g_{tt}(r')| P_s(r')  4 \pi r'^2 \sqrt{g_{rr}(r')}dr'=\frac{1}{r},
\end{eqnarray}
which is the same result as in the in the Newtonian problem. However, it is interesting that only a fraction of the radiation emitted at any given $r$ escapes to infinity and can be thus observed. It is known that the solid angle of the
escaping rays is equal to $2\pi(1 +\cos\theta)$ for $r$ outside the photon sphere radius, i.e. $r>r_{ph}$ and
$2\pi(1 -\cos\theta)$ for $r$ smaller then the photon radius, ie. $r<r_{ph}$. Using the fact that the critical
rays that barely escape have specific angular momentum equal to
that of the photon orbit one can show the relation
\begin{eqnarray}
\sin\theta= \frac{b}{r} \sqrt{1-\frac{2 \mathcal{M}(r)}{r}},
\end{eqnarray}
where $b$ is the impact parameter of the photon orbit. In the case of a black hole spacetime surrounded by dark matter for the escaping rays, the net luminosity observed is found to be  \cite{Narayan}
\begin{equation}
L^{DM}=\int_{r_{ph}}^{r} j_s(r')\, 2\pi \left(1 +\sqrt{1-\frac{b^2}{r'^2} \,f(r')}\right)4 \pi r'^2 dr'.
\end{equation}
For model I, using the parameters $M=m=1$, $r_s=2 m$, and $\Delta r_s=10 M$, we find 
\begin{eqnarray}
L^{DM}=0.347126.
\end{eqnarray}
In this example, the observer is located at the distance $r_{obs.}=10^6 m$. Consider now a more realistic situation when the observer is located within the dark matter halo using the parameters $m=1$, $M=10^3m$, $r_s=2 m$, and $\Delta r_s=10^6 M$, and say the observer is located at the distance $r_{obs.}=10^3m$. In that case, we find 
\begin{eqnarray}
L^{DM}=0.335963.
\end{eqnarray}
We see that when the observer is located inside the dark matter halo, the net luminosity decreases. 
In the perfect fluid dark matter model using (18), the spacetime is asymptotically flat, hence the net luminosity observed at infinity using the parameters $m=1$ along with $a=0.1$, and $a=0.2$ is found to be 
\begin{eqnarray}
L^{DM}_{\infty}=0.388892,
\end{eqnarray}
and
\begin{eqnarray}
L^{DM}_{\infty}=0.416716,
\end{eqnarray}
respectively. In this case we see that the net luminosity observed at infinity is considerably higher compared to the Schwarzschild vacuum solution. Moreover, this strong effect shows that metric function (18) describes the spacetime near the black hole surrounded by dark matter with relatively high energy density.  The increase in luminosity shows that, the gravitational deflection of light rays when the perfect fluid dark matter is present, is considerably smaller as compared to the Schwarzschild vacuum black hole. In other words, in this case,  a larger fraction of the emitted radiation at any $r$ escapes from the black hole. As a result, we observe a higher value of the net luminosity.
Finally, using (30) we can find the net luminosity observed at the distance $r_{obs}=r_0=10^6$ using the parameters $m=1$ along with $\alpha=\gamma=0.1$, given by
\begin{eqnarray}
L^{DM}=0.339701.
\end{eqnarray}

Compared to the Schwarzschild vacuum solution, we observe a very small increase in the net luminosity. This shows that the gravitational deflection angle of light in this dark matter model is slightly decreased compared to the Schwarzschild vacuum black hole. We can say that the metric function (30) describes the spacetime around black hole with lower concentration of dark matter while the metric function (18) describes the dark matter effect in the galactic center where the effect is stronger. For example, one can use (30) to explain the flat curve for the observed motions of stars in the outer part of galaxies, while we can use the metric function (18) to describe the motion of stars in the galactic center. Note that in all cases, there is a tiny contribution in the intensity plots even for the region $r<r_{ph}$. In fact, as we are going to see, the intensity even inside the shadow does not go to zero but has a small finite value. Therefore, if we  integrate from the horizon to the photon sphere we end up with a very tiny effect in the luminosity which is  included in the shadow/intensity plots.
Next, to determine the image observed by a distant observer, we can use the observed specific intensity at the observed photon frequency which can be found by integrating the specific emissivity along the photon path
\begin{eqnarray}
    I_{obs}(\nu_{obs}) = \int_{\gamma}\mathrm{g}^3 j(\nu_{e})dl_\text{prop},  
\end{eqnarray}
where $\mathrm{g} = \nu_{obs}/\nu_{e}$ is the redshift factor, $\nu_{e}$ is the photon frequency as measured in the rest-frame of the emitter, $j(\nu_{e})$ is the emissivity per unit volume in the rest-frame of the emitter, while $dl_\text{prop} = k_{\alpha}u^{\alpha}_{e} d\lambda$ is the infinitesimal proper length as measured in the rest-frame of the emitter. The redshift factor is evaluated from \cite{Bambi:2013nla}
\begin{eqnarray}
\centering
    \mathrm{g} = \frac{k_{\alpha}u^{\alpha}_{\text{obs}}}{k_{\beta}u^{\beta}_{e}}  
\end{eqnarray}
where $k^{\mu}$ is the four-velocity of the photons, $u^{\alpha}_{e}$ four-velocity of the accreting gas emitting the radiation, $u^{\mu}_{\text{obs}}$ = $(1,0,0,0)$ represents the  four-velocity of the distant observer, and finally $\lambda$ is the affine parameter along the photon path $\gamma$. Here $\gamma$ in the integral refers that the integral has to be evaluated along the path of the photon (null geodesics).
Using the rest spherical accretion model, the redshift factor is evaluated from $\mathrm{g}=f(r)^{1/2}$. For the specific emissivity we assume a simple model in which the emission is monochromatic with emitter's-rest frame frequency $\nu_{\star}$, and the emission has a $1/r^2$ radial profile:
\begin{eqnarray}
    j(\nu_{e}) \propto \frac{\delta(\nu_{e}-\nu_{\star})}{r^2},
\end{eqnarray}
where $\delta$ is the Dirac delta function. The proper length can be written as 
\begin{eqnarray}
    dl_{\text{prop}} =\sqrt{f(r)^{-1}dr^2+r^2d \phi^2}.
\end{eqnarray}

If follows that the specific intensity observed by the
infinite observer is
\begin{equation}
I_{obs}(\nu_{obs}) = \int_{\gamma} \frac{f(r)^{3/2}}{r^2} \sqrt{f(r)^{-1}+r^2 (\frac{d\phi}{dr})^2}  dr 
\end{equation}
where 
\begin{eqnarray}
\frac{dr}{d\phi}=\pm r^2 \sqrt{\frac{1}{b^2}-\frac{f(r)}{r^2}},
\end{eqnarray}
 is obtained from the equation of motion obtained for the light ray that moves on the equatorial plane.
 

\section{Optically thin radiating and infalling gas surrounding a black hole}
In this section, we consider a more realistic model, namely an optically thin, radiating accretion flow surrounding the object and then use a numerical technique (Backward Raytracing) to find the shadow cast by the radiating flow. The calculation of the intensity map of the emitting region requires some assumption about the radiating processes and emission mechanisms. The observed specific intensity $I_{\nu 0}$ at the observed photon frequency $\nu_\text{obs}$ at the point $(X,Y)$ of the observer's image (usually measured in $\text{erg} \text{s}^{-1} \text{cm}^{-2} \text{str}^{-1} \text{Hz}^{-1}$) is given by \cite{Bambi:2013nla}
\begin{eqnarray}
    I_{obs}(\nu_{obs},X,Y) = \int_{\gamma}\mathrm{g}^3 j(\nu_{e})dl_\text{prop},  
\end{eqnarray}
 
Here we are considering a simplistic case of the accreting gas. We assume that the gas is in radial free fall with a four-velocity which in a static and spherically symmetric case reduces to 
\begin{eqnarray}
u^t_{e} & = & \frac{1}{f(r)}, \nonumber \\
u^r_{e} & = & -\sqrt{\frac{g(r)}{f(r)}\left(1-f(r)\right)}, \nonumber \\
u^{\theta}_{e} & = & 0, \nonumber \\
u^{\phi}_{e} & = & 0, 
\end{eqnarray}
where 
\begin{eqnarray}
f(r)=g(r)=1-\frac{2\mathcal{M}(r)}{r}.
\end{eqnarray}
 The four-velocity for the photons has been found in the previous section. To ease our further calculations, we find a relation between the radial and time component of the four-velocity
\begin{eqnarray}
    \frac{k^r}{k^t} = \pm f(r) \sqrt{g(r)\bigg(\frac{1}{f(r)}-\frac{b^2}{r^2}\bigg)},
\end{eqnarray}
where the sign +(-) is when the photon approaches (goes)
away from the massive object.
The redshift function $\mathrm{g}$ is therefore given by
\begin{eqnarray}
  \mathrm{g} = \frac{1}{\frac{1}{f(r)} \pm \frac{k_r}{k_t}\sqrt{\frac{g(r)}{f(r)}\bigg(1-f(r)\bigg)}}.
\end{eqnarray}
For the specific emissivity we assume a simple model in which the emission is monochromatic with emitter's-rest frame frequency $\nu_{\star}$, and the emission has a $1/r^2$ radial profile:
\begin{eqnarray}
    j(\nu_{e}) \propto \frac{\delta(\nu_{e}-\nu_{\star})}{r^2},
\end{eqnarray}
where $\delta$ is the Dirac delta function. The proper length can be written as 
\begin{eqnarray}
    dl_{\text{prop}} = k_{\alpha}u^{\alpha}_{e}d\lambda = -\frac{k_t}{\mathrm{g}|k^r|}dr,
\end{eqnarray}
Integrating the intensity over all the observed frequencies, we obtain the observed flux
\begin{eqnarray} \label{inten}
    F_{obs}(X,Y) \propto -\int_{\gamma} \frac{\mathrm{g}^3 k_t}{r^2k^r}dr.  
\end{eqnarray}

In Fig. 2 we show the shadow images and the corresponding intensities using the rest gas model as seen by a distant observer for the Schwarzschild black hole surrounded by dark matter using the mass function (1). We see that having a high density of dark matter near the black hole affects the images as well as the intensity. For instance, it can be observed from Fig. 2 that for a given set of values of the parameters, the shadow size is decreased or increased in comparison to the Schwarzschild vacuum black hole. This effect in principle can be used as a tool to observe very dense dark matter located around black holes. On the other hand, in Fig. 3 we show shadow images and the intensities using the infalling gas model as seen by a distant observer for the black hole surrounded by dark matter using the mass function (1). Again we cam observe that for a given domain of parameters for the dark matter we can distinguish such a black hole compared to the Schwarzschild vacuum black hole.  Finally, in Figs. 4-5 we show the shadow images and intensities for the black hole surrounded by perfect fluid dark matter using the mass function (18) and (30), respectively.  It's worth noting that, the perfect fluid dark matter model is more realistic model compared to the Model I which is just a toy model to describe surrounding matter around black hole (not necessarily dark matter).  From the Figs. 4-5 it can be seen that the particular model of dark matter has a remarkable role on the final image. In other words, depending on the particular gas model as well as the dark matter model, the shadow images and the intensities are different from the Schwarzschild vacuum black hole. This means that, in principle, shadow images can be used as a tool to observe dark matter around black holes.  It is interesting to see how the contribution of the intensity of static gas model obtained from the region inside the shadow region (part of the radiation that has escaped to infinity) is higher compared to the infalling gas model. The closer it the dark matter to the black hole the stronger is the effect on the shadow images and and this has to do with the fact that the photon radius is changed due to the surrounding matter.  

\subsubsection{Backward Ray tracing} \label{ray}
Now, when we have the desired relation for the observed flux, we are ready to run our simulation. The technique that is used for making shadows of compact objects is Relativistic Ray tracing. There are two different methods which one can be used to do ray tracing; forward ray tracing and backward ray tracing. In forward ray tracing, the light emitted from the source, gets lensed and, red/blue-shifted from the object and then reaches the observer's eye. Here we follow the path of the photon from the point of emission/source to the observer's eye. In terms of computing, we replace the eye with an image plane made of pixels, so when a photon hits one of those many pixels, it gets lightened up indicating that the value of the pixel at that point is greater than zero. This process is repeated multiple times till the image plane fills up and is properly adjusted. The limitation of this technique is, we assume that the rays always hit the image plane, whereas in the real scenario the photons have a very small probability of actually hitting the eye. Potentially, we'll have to shoot millions and billions of rays to get only a single photon actually reaching the eye. One can argue that instead of shooting in all random directions, just shoot where the eye is. This may optimise the problem to some extent, but it is not important for the direction to be specific for diffused surfaces, which is a matter of light-matter interaction. To sum up, this increases the computational power, which means the integration of the  rays will neither be that faster nor efficient. Therefore we move on to the second method which is 'backward ray tracing'. Here, instead of tracing the ray from the source towards the eye, we trace the rays backwards in time, which means from the eye towards the source.
\begin{table*}[tbp] 
        \begin{tabular}{|l|l|l|l|l|l|}
        \hline
    \multicolumn{1}{|c|}{ } &  \multicolumn{1}{c|}{  $l=1$ } 
    & \multicolumn{1}{c|}{  $l=2 $} &   \multicolumn{1}{c|}{  $l=3$ } 
    & \multicolumn{1}{c|}{  $l=4$ } & \multicolumn{1}{c|}{} \\
    \hline
  $\Delta r_s [m]$ & \quad $\omega_{\Re}$ & \quad $\omega_{\Re}$ & \quad $\omega_{\Re}$  
  & \quad $\omega_{\Re}$   & \quad $R_s$  \\ 
        \hline 
10 & 0.280776 &  0.46796 & 0.655144 & 0.842328 & 5.34234  \\
$10^2$ & 0.288589 &  0.480982 & 0.673375  & 0.865767  & 5.19770 \\
$10^3$ & 0.288674 &  0.481124 &  0.673573 &    0.866023 & 5.19617 \\
$10^6$ & 0.288675 & 0.481125 &  0.673575 & 0.866025  & 5.19615 \\
        \hline
        \end{tabular}
         \caption{ \label{table4} Numerical values of the shadow radius and the 
    real part of QNMs frequencies obtained for the black hole surrounded by dark matter using  $m=1$, $r_s=2 m$ and positive energy density $M=1$. }
\end{table*}

\begin{table*}[tbp] 
        \begin{tabular}{|l|l|l|l|l|l|}
        \hline
    \multicolumn{1}{|c|}{ } &  \multicolumn{1}{c|}{  $l=1$ } 
    & \multicolumn{1}{c|}{  $l=2 $} &   \multicolumn{1}{c|}{  $l=3$ } 
    & \multicolumn{1}{c|}{  $l=4$ } & \multicolumn{1}{c|}{} \\
    \hline
  $\Delta r_s [m]$ & \quad $\omega_{\Re}$ & \quad $\omega_{\Re}$ & \quad $\omega_{\Re}$  
  & \quad $\omega_{\Re}$   & \quad $R_s$  \\ 
        \hline 
10 & 0.296932 &  0.494887 & 0.692842 & 0.890797 & 5.05166 \\
$10^2$ & 0.288761 &  0.481269 & 0.673776  & 0.866284  & 5.19460 \\
$10^3$ & 0.288676 &  0.481127 &  0.673577 & 0.866028  & 5.19614 \\
$10^6$ & 0.288675 & 0.481125 & 0.673575  & 0.866025 & 5.19615 \\
        \hline
        \end{tabular}
         \caption{ \label{table4} Numerical values of the shadow radius and the 
    real part of QNMs frequencies obtained for the black hole surrounded by dark matter using  $m=1$, $r_s=2 m$ and negative energy density having $M=-1$.}
\end{table*}
\begin{table*}[tbp]
        \begin{tabular}{|l|l|l|l|l|l|}
        \hline
    \multicolumn{1}{|c|}{ } &  \multicolumn{1}{c|}{  $l=1$ } 
    & \multicolumn{1}{c|}{  $l=2$ } &   \multicolumn{1}{c|}{  $l=3$ } 
    & \multicolumn{1}{c|}{  $l=4$ } & \multicolumn{1}{c|}{} \\
    \hline
  $a [m]$ & \quad $\omega_{\Re}$ & \quad $\omega_{\Re}$ & \quad $\omega_{\Re}$  
  & \quad $\omega_{\Re}$   & \quad $R_s$  \\ 
        \hline 
0.1 & 0.334965 & 0.558275 & 0.781585  & 1.0049 & 4.35581 \\
0.2 & 0.359565 & 0.599274 &  0.838984 &  1.07869 & 3.92406  \\
0.3  & 0.376722 & 0.62787 &  0.879019 &  1.13017  & 3.65108 \\
0.4  & 0.389437 & 0.649062 &  0.908687  &  1.16831 & 3.47439 \\
        \hline
        \end{tabular}
         \caption{ \label{table4} Numerical values of the shadow radius and the 
    real part of QNMs frequencies obtained for the black hole surrounded by perfect fluid dark matter as a function of $a$.}
\end{table*}
\begin{table*}[tbp]
        \begin{tabular}{|l|l|l|l|l|l|}
        \hline
    \multicolumn{1}{|c|}{ } &  \multicolumn{1}{c|}{  $l=1$ } 
    & \multicolumn{1}{c|}{  $l=2$ } &   \multicolumn{1}{c|}{  $l=3$ } 
    & \multicolumn{1}{c|}{  $l=4$ } & \multicolumn{1}{c|}{} \\
    \hline
  $\alpha$ & \quad $\omega_{\Re}$ & \quad $\omega_{\Re}$ & \quad $\omega_{\Re}$  
  & \quad $\omega_{\Re}$   & \quad $R_s$  \\ 
        \hline 
$0.1$ & 0.286865 & 0.478108 & 0.669352 & 0.860595  &  5.22894 \\
$0.2$ & 0.278488 & 0.464146 & 0.649804  & 0.835463  &  5.38624 \\
$0.3$  & 0.276151 &  0.460251 &  0.644352 &  0.828452 & 5.43182 \\
$0.4$  & 0.275496 & 0.459160 &  0.642824  & 0.826488  &  5.44473 \\
        \hline
        \end{tabular}
         \caption{ \label{table4} Numerical values of the shadow radius and the 
    real part of QNMs frequencies obtained for the black hole surrounded by perfect fluid dark matter as a function of $\alpha$. Here we chose $\gamma=0.1$ and $r_0=10^6$. }
\end{table*}
To implement a ray tracer, the following steps are performed:
\begin{enumerate}
    \item Multiple parallel beams of light with varying impact parameters are shot from the observer at $r_0 = 30M$ towards the object.
    
    \item Due to a defined critical impact parameter, the rays which have $b>b_{crit}$ get deflected and reach the source, while the ones with $b<b_{crit}$ plunge into the the singularity.
    
    \item Rays that get deflected travel from the observer towards the emitter and possess a certain turning point $r_{tp}$. So from the observer till $r_{tp}$ the light turns out to be red-shifted and from $r_{tp}$ till the emitter, it turns out to be blue-shifted and illuminates the region. The impact parameter is related to the $r_{tp}$ of a photon ($\dot{r}=0, V_{\text{eff}}=0$):
    \begin{equation}
        b = \frac{r_{tp}}{\sqrt{f(r_{tp})}}
    \end{equation} 
    
    \item Rays that travel below the $b_{crit}$ have no turning points in turn going into the singularity and have only a high red-shifted shadow, which darkens the region, hence giving out the shadow.
    \item Intensity is then calculated by integrating Eq. (\ref{inten}) from the observer to the emission point $(r_{em})$.
    \begin{equation}
        F_{obs}(X,Y) = -\int_{r_{obs}}^{r_{tp}} \frac{\mathrm{g}^3 k_t}{r^2k^r}dr - \int_{r_{tp}}^{r_{em}} \frac{\mathrm{g}^3 k_t}{r^2k^r}dr.  
    \end{equation}
    
We created a routine in python following the above steps mentioned for ray-tracing. A simplified version of the code using the Schwarzschild spacetime has been previously implemented in the \texttt{EinsteinPY} library \cite{Epy1, Bapat:2020xfa} which was subsequently used in \cite{Kala:2020prt} as well.
\end{enumerate}

\begin{figure*}
   	\includegraphics[width=7 cm]{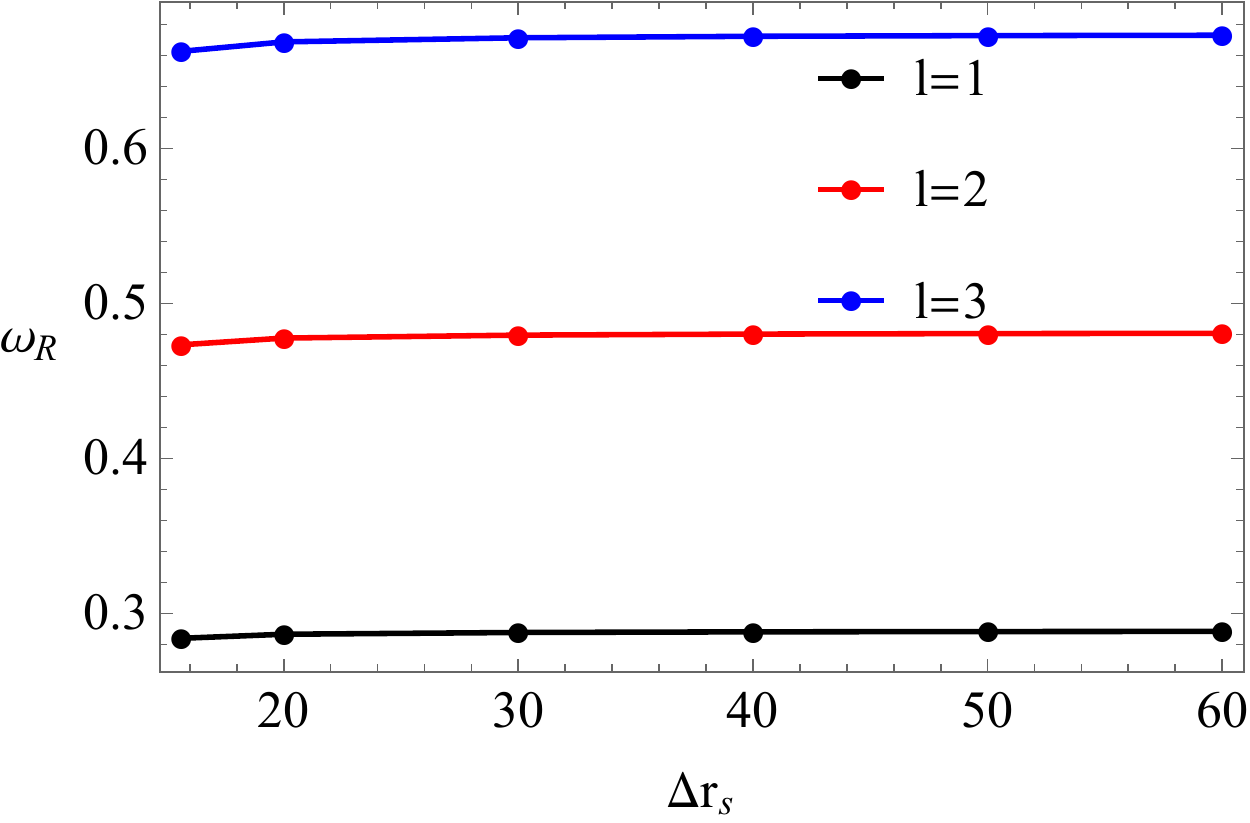}
 \includegraphics[width=7 cm]{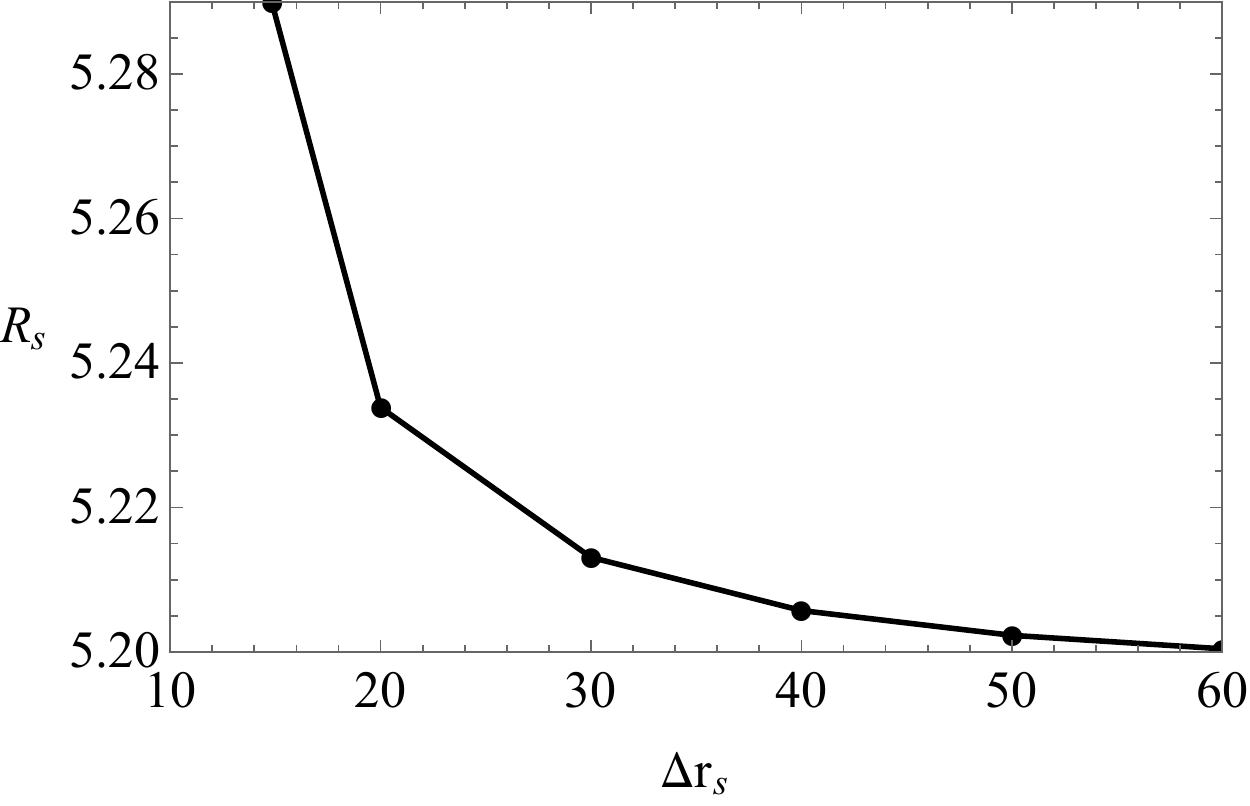}
    \caption{QNMs frequencies obtained for the black hole surrounded by dark matter using  $m=1$, $r_s=2 m$ and positive energy density having $M=1$.  \label{rs}}
\end{figure*}
\begin{figure*}
   	\includegraphics[width=7 cm]{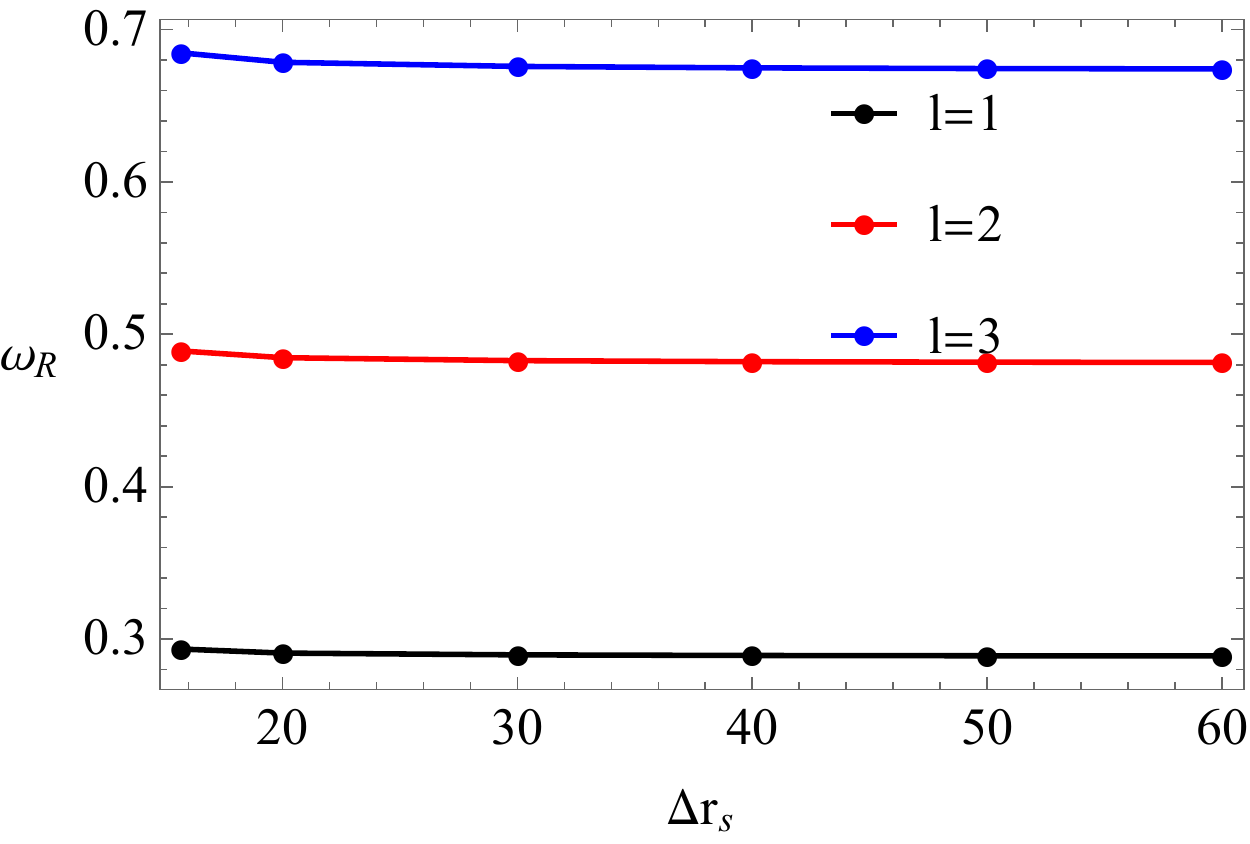}
 \includegraphics[width=7 cm]{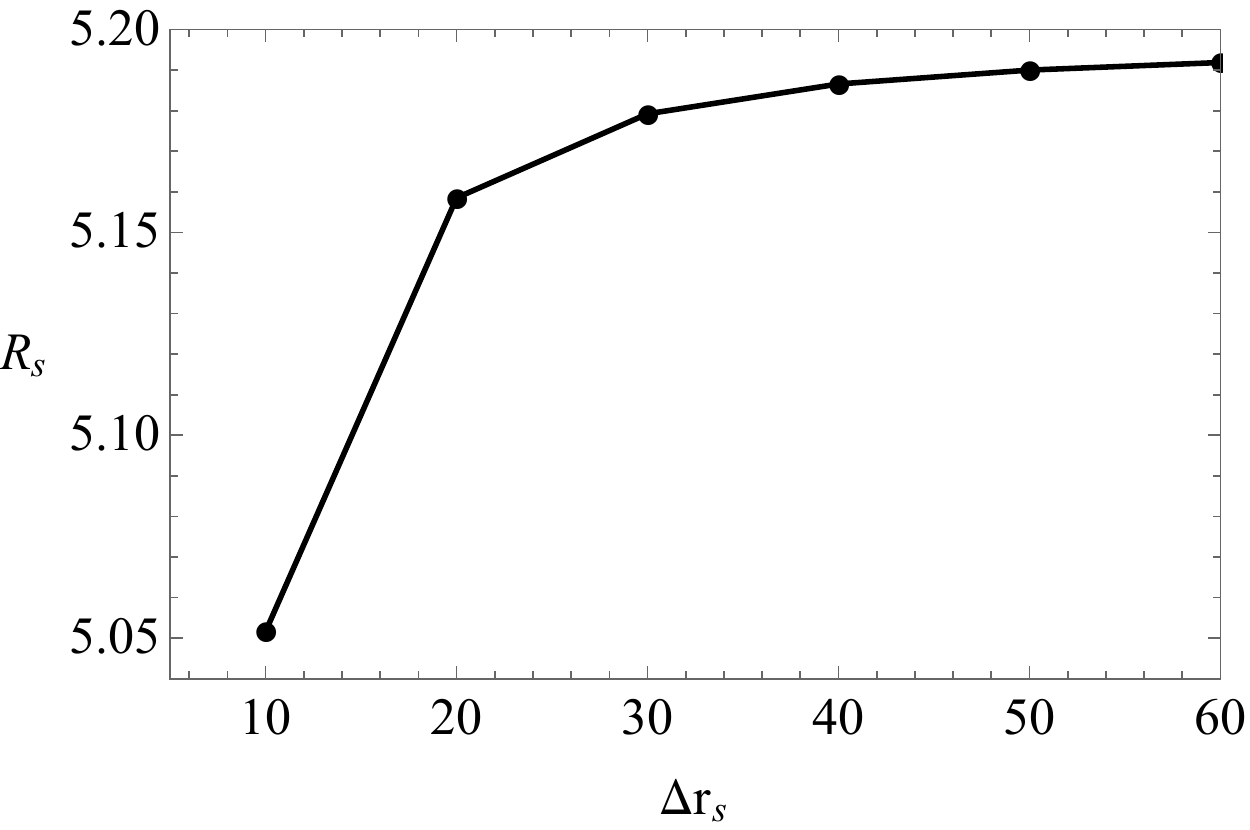}
    \caption{QNMs frequencies obtained for the black hole surrounded by dark matter using $m=1$, $r_s=2 m$ and negative energy density having $M=-1$.  \label{rs}}
\end{figure*}
\begin{figure*}
   	\includegraphics[width=7 cm]{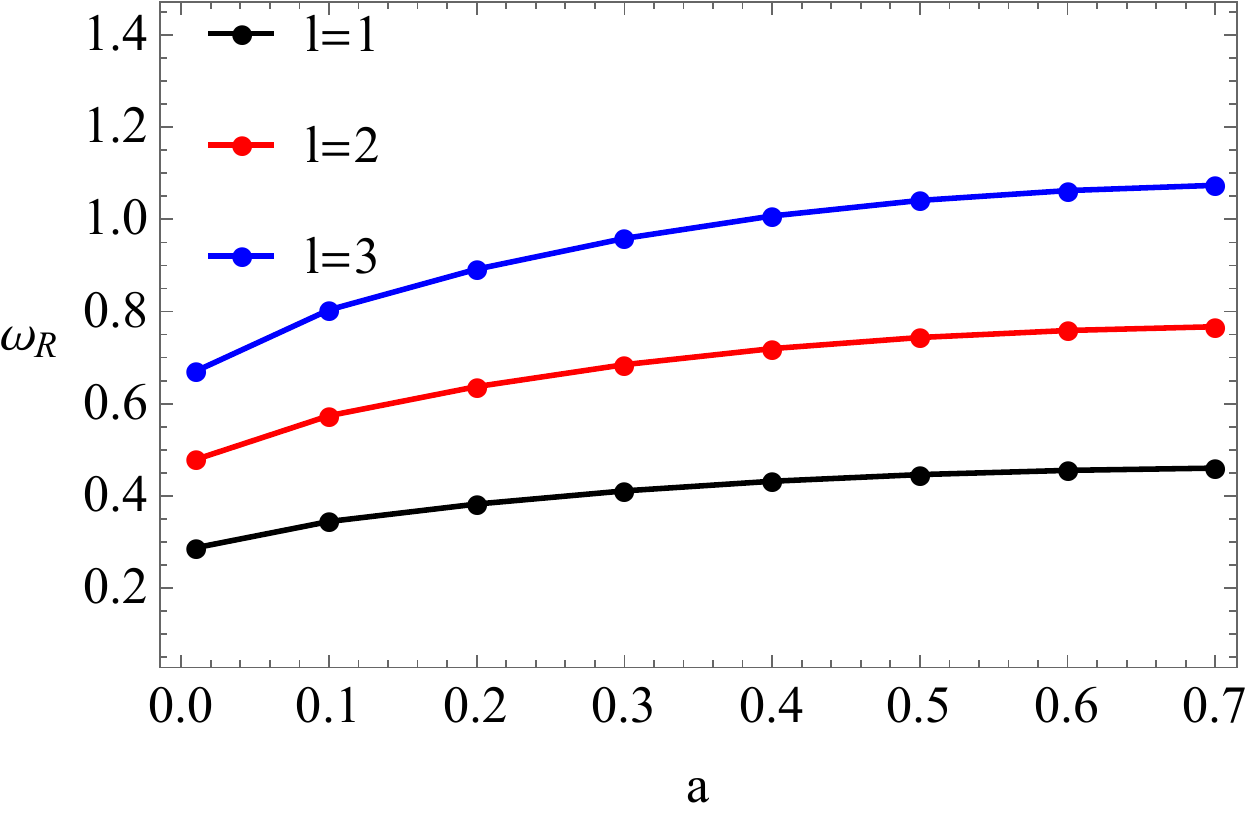}
 \includegraphics[width=7 cm]{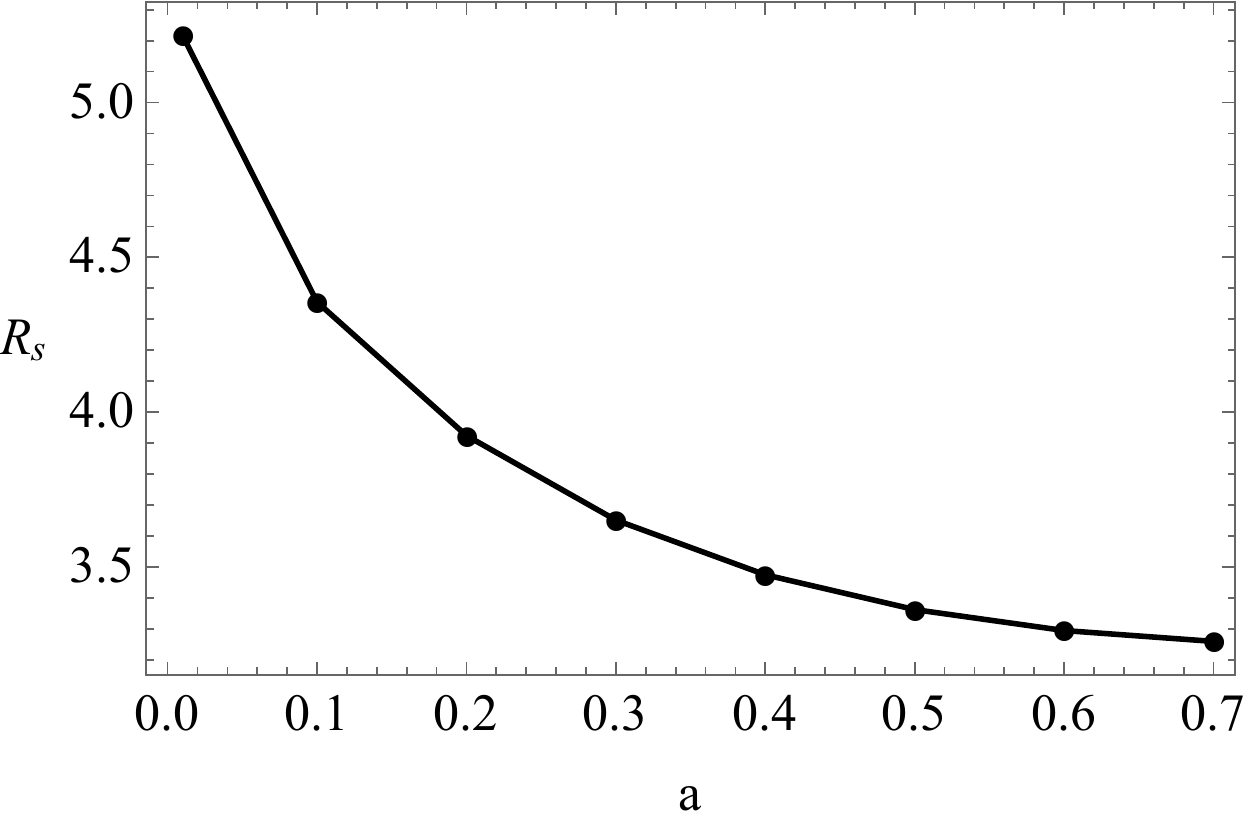}
    \caption{Shadow radius and the real part of QNMs as  a function of the parameter $a$.  \label{rs}}
\end{figure*}
\begin{figure*}
   	\includegraphics[width=7 cm]{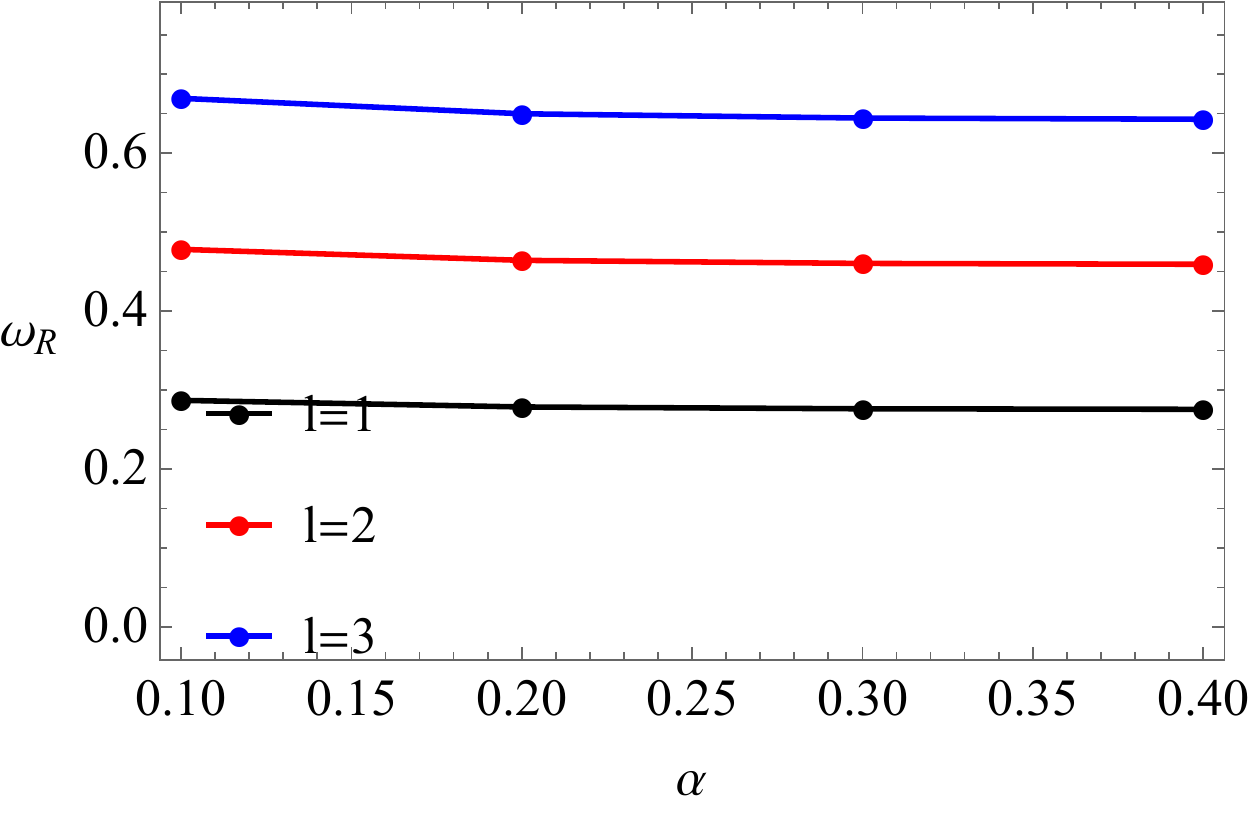}
 \includegraphics[width=7 cm]{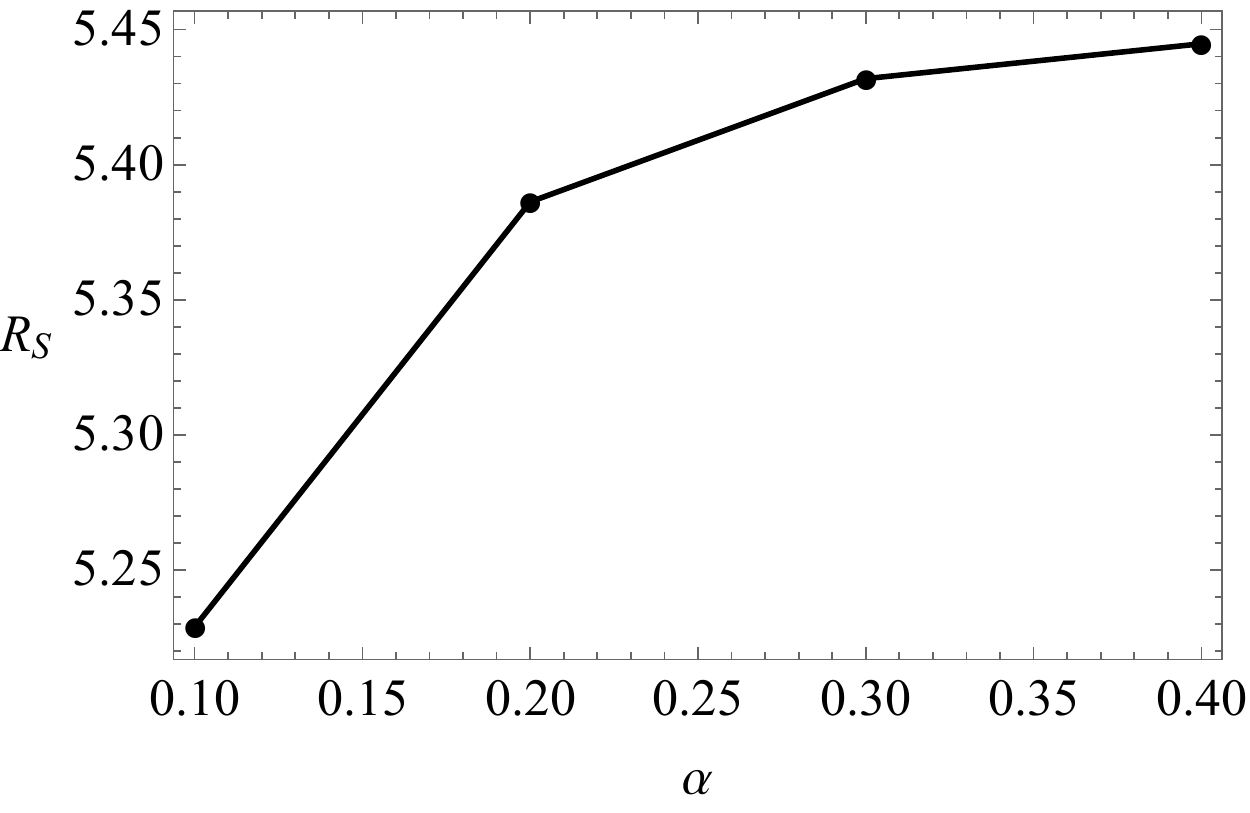}
    \caption{Shadow radius and the real part of QNMs as  a function of the parameter $\alpha$. We have $r_0=10^6$ and $\gamma=0.1$.  \label{rs}}
\end{figure*}

\section{Correspondence between the shadow radius and QNMs in dark matter spacetime}
\label{QNMs-shad}
In this final section we proceed to investigate another possible effect in the spacetime of dark matter based on the field perturbations. In particular we shall investigate the correspondence between the black hole shadow in dark matter halo and 
the real part of QNMs frequencies. The perturbation of the black hole during the ring down phase can be studied in terms of the QNMs where one must impose an outgoing boundary condition at infinity and an ingoing boundary condition at the horizon. One can show that QNMs can be written in terms of the real part and the imaginary part representing the decaying modes
\begin{equation}
\omega_{QNM}=\omega_{\Re}-i \omega_{\Im}.
\end{equation}
Previously, it has been shown that in the eikonal limit, the real part of the 
QNMs frequencies is mathematically related to the angular velocity of the unstable 
null geodesic \cite{cardoso,Wei:2019jve,Stefanov:2010xz}. Furthermore, the imaginary part of the QNMs frequencies is  linked with the Lyapunov exponent that determines the instability time
scale of the orbits \cite{cardoso} 
\begin{equation}
	\omega_{QNM}=\Omega_c l -i \left(n+\frac{1}{2}\right)|\lambda|, 
	\label{41}
\end{equation}
where $l$ is the multipole number, $\Omega_c$ is the angular velocity at the unstable null geodesic, and $\lambda$ denotes the Lyapunov exponent. In recent works, it was argued that the real part of QNMs frequencies and shadow radius (if we also include a sub-leading regime to half 
of its value) are related as follows
(see for details \cite{Jusufi:2019ltj,Cuadros-Melgar:2020kqn})
\begin{equation}\label{modeq}
    \omega_{\Re}=R_s^{-1}\left(l+\frac{1}{2}\right),
\end{equation}
in which $R_s$ is the radius of the black 
hole shadow. This equation is accurate in the eikonal limit having large values of 
multipole number, i.e., $l>>1$. In that case we can further simplify if as $  \omega_{\Re}=R_s^{-1}l$.
It is thus interesting to see if we can use the above correspondence to calculate the QNMs  frequencies once we have the black hole shadow radius. As we explained, this geometric-optics correspondence  becomes accurate only for large $l$, but sometimes works well even for the fundamental modes (compared to the scalar field perturbations). Given the fact that fundamental modes are more important from observational point of view, we present our numerical results of the real part for some fundamental QNMs 
frequencies presented in Tables I-IV. From Table I we observe that the shadow radius decreases with the decrease in dark matter density (increasing $\Delta r_s)$. From the inverse relation between the shadow radius and the QNMs frequencies, this means that the real part of QNMs should increase with the decrease in $\Delta r_s$. In fact, we can verify this fact from the numerical values of QNMs given in Table I. In Table II, we consider the case of having dark matter with negative energy density.  Quite interestingly, in this case we find that the values of the shadow radius are smaller compared to the Schwarzschild vacuum solution and increase with the decrease of energy density. The QNMs frequencies decrease with the decrease of the of energy density.  This suggest that the black hole should oscillate more rapidly if we increase the amount of negative energy density of dark matter. A similar results are obtained for the perfect fluid dark matter with the metric function (18) and (30), respectively. In both cases the  shadow radius is smaller compared to the Schwarzschild vacuum solution, however in the first case the shadow decreases with $a$ and increases with $\alpha$ in the latter case.  The fact that the shadow radius is smaller as compared to the Schwarzschild vacuum black hole is explained from the fact that dark matter with negative energy density is involved in the perfect fluid model, provided $a>0$ and $\alpha>0$, respectively. For more details we refer the reader to the plots of QNMs frequencies and the shadow radius shown in Figs. 6-9. Note that the fundamental modes of QNMs and the shadow radius in a static perfect fluid dark matter without accretion matter with the metric function (18) have been investigated in Ref. [68] using the WKB method. Our results in Table III, are in good agreement with the scalar field perturbations presented in [68].  \\

\section{Conclusions}
We have studied the influence of dark matter on the shadow images using the electromagnetic radiations emitted from spherical accretion medium which was assumed to be an optically-thin region surrounding the black hole.  We have considered two spherical accretion models: optically-thin radiating gas at rest and a  gas in a radial free fall around the static and spherically symmetric black hole. We have shown that due to the effect of dark matter on the spacetime geometry the intensity of the electromagnetic flux radiation is altered compared to the Schwarzschild vacuum black hole.  In particular we have investigated two models for the dark matter distribution: Firstly, the case of total mass function which consists a black hole  with mass $m$, surrounded by the dark matter having mass $M$ and positive energy density along with a thickness $\Delta r_s$ has been analyzed. Secondly, we explore the case having exotic dark matter around the black hole with negative density energy and a thickness $\Delta r_s$.\\

In order to have a more realistic model we need to take into account the relativistic pressure of dark matter, therefore we can consider a perfect fluid dark matter surrounding the black hole.  It is found that in order to have significant effect of dark matter on the intensity of the electromagnetic flux radiation, a high energy density of dark matter near the black hole is needed.  Among other things, we have shown the surrounding dark matter mass function which describes the distribution of the dark matter plays a key role on the effect of shadow images. Furthermore, we argue that if the surrounding dark matter has positive (negative) energy density, the shadow radius and the intensity of the electromagnetic flux radiation increases (decreases), respectively. Finally, we have also studied the  correspondence between the QNMs frequencies and the black hole shadow radius. We have shown that the shadow radius as well as the QNMs frequencies can decrease or increase depending on the concentration of dark matter near the black hole having positive or negative energy density. \\

We also note that the observed image will not depend on the distance
units being used, as long as the distance between the observer and the
black hole is kept the same. However, a more delicate issue is a comparison
between images of different black holes. In such a case, saying that the horizon
mass $m$ is the same on both systems or that the ADM mass $M_{ADM}$ is the same
on both systems will not be the equivalent statements in general. This means that, the
conclusion that the shadow image is larger in one system than another
strongly depends on which mass scale is set to be the same on both
systems. \\

Finally, we should mention here that the gas medium in the present paper is optically thin accreting medium, we plan in the near future to extend our result by including a rotating accretion disk and a rotating black hole spacetime. This is of course outside the scope of this article. \\


\begin{thebibliography}{99}


\bibitem{Cunha:2020azh}
P.~Cunha, V.P. and C.~A.~R.~Herdeiro,
Phys. Rev. Lett. \textbf{124}, no.18, 181101 (2020)

\bibitem{Akiyama1} K. Akiyama et al. (Event Horizon Telescope), Astrophys.
J. 875 (2019) L1.

\bibitem{Akiyama4} K. Akiyama et al. (Event Horizon Telescope), Astrophys.
J. 875 (2019) L4.

\bibitem{AbbottBH} B. P. Abbott et al. (LIGO Scientific and Virgo
Collaborations), Phys. Rev. Lett. 116 (2016) 061102.


\bibitem{Do:2019txf}
T.~Do, A.~Hees, A.~Ghez, G.~D.~Martinez, D.~S.~Chu, S.~Jia, S.~Sakai, J.~R.~Lu, A.~K.~Gautam and K.~K.~O'Neil, \textit{et al.}
Science \textbf{365}, no.6454, 664-668 (2019)

\bibitem{Naoz:2019sjx}
S.~Naoz, C.~M.~Will, E.~Ramirez-Ruiz, A.~Hees, A.~M.~Ghez and T.~Do,
Astrophys. J. Lett. \textbf{888}, no.1, L8 (2020)

\bibitem{Do:2008uf}
T.~Do, A.~M.~Ghez, M.~R.~Morris, S.~Yelda, L.~Meyer, J.~R.~Lu, S.~D.~Hornstein and K.~Matthews,
Astrophys. J. \textbf{691}, 1021-1034 (2009)


  \bibitem{Synge66} 
  J.~L.~Synge,
  Mon.\ Not.\ Roy.\ Astron.\ Soc.\  {\bf 131}, no. 3, 463 (1966).
 
   
  \bibitem{Luminet79} 
  J.-P.~Luminet,
  Astron.\ Astrophys.\  {\bf 75}, 228 (1979).

  \bibitem{DeWitt73} 
  J. M. Bardeen, in Black Holes (Proceedings, Ecole d'Eté de Physique Théorique: Les Astres Occlus : Les Houches, France, August, 1972) edited by C.~DeWitt and B.~S.~DeWitt
  

 \bibitem{cardoso} V. Cardoso, A. S. Miranda, E. Berti, H. Witek, and V. T. Zanchin, Phys. Rev. D 79, 064016 (2009)

\bibitem{Wei:2019jve}
  S.~W.~Wei and Y.~X.~Liu,
  arXiv:1909.11911 [gr-qc].
  
  
  \bibitem{Stefanov:2010xz}
  I.~Z.~Stefanov, S.~S.~Yazadjiev and G.~G.~Gyulchev,
  Phys.\ Rev.\ Lett.\  {\bf 104} (2010) 251103
  
 \bibitem{Konoplya:2017wot}
  R.~A.~Konoplya and Z.~Stuchlík,
  Phys.\ Lett.\ B {\bf 771} (2017) 597
  
\bibitem{Liu:2020ola}
  C.~Liu, T.~Zhu, Q.~Wu, K.~Jusufi, M.~Jamil, M.~Azreg-Aïnou and A.~Wang,
  Phys.\ Rev.\ D {\bf 101} (2020) no.8,  084001
  
\bibitem{Feng:2019zzn}
  X.~H.~Feng and H.~Lu,
  arXiv:1911.12368 [gr-qc].
\bibitem{Dey:2020a}
  D. Dey, R. Shaikh and P.S. Joshi,
  arXiv:2009.07487 [gr-qc].
 
\bibitem{Dey:2020b}
  A.B. Joshi, D. Dey, P.S. Joshi and P. Bambhaniya,
  arXiv:2004.06525 [gr-qc].


\bibitem{Dey:2020c}
    D. Dey, R. Shaikh and P.S. Joshi,
  arXiv:2004.06525 [gr-qc].  
  
\bibitem{Zhang:2019glo}
  M.~Zhang and M.~Guo,
  arXiv:1909.07033 [gr-qc].
  
  
 \bibitem{000} 
  K.~Hioki and K.~i.~Maeda,
  Phys.\ Rev.\ D {\bf 80} (2009) 024042
  
\bibitem{00} 
 S.-W.~Wei, Y.-X.~Liu, R. B.~Mann,
Phys. Rev. D {\bf 99}, 041303 (2019).

\bibitem{01} 
 S.-W.~Wei, Y.-C.~Zou, Y.-X.~Liu, R. B.~Mann,
JCAP {\bf 1908}, 030 (2019).

 
\bibitem{22}
  T.~Zhu, Q.~Wu, M.~Jamil and K.~Jusufi,
  Phys.\ Rev.\ D {\bf 100} (2019) no.4,  044055.
  
\bibitem{55}
K. Jusufi, M. Jamil, H. Chakrabarty, Q. Wu, C. Bambi, A. Wang,
Phys. Rev. D \textbf{101}, 044035 (2020).


 \bibitem{66} 
 C.~Bambi and K.~Freese,
 Phys.\ Rev.\ D {\bf 79}, 043002 (2009).
 
 \bibitem{77} 
  C.~Bambi and N.~Yoshida,
  Class.\ Quant.\ Grav.\  {\bf 27}, 205006 (2010).
 
 
 \bibitem{88} 
 A.~Abdujabbarov, M.~Amir, B.~Ahmedov and S.~G.~Ghosh,
 Phys.\ Rev.\ D {\bf 93}, no. 10, 104004 (2016).
 
 \bibitem{99} 
 M.~Amir and S.~G.~Ghosh,
 Phys.\ Rev.\ D {\bf 94}, no. 2, 024054 (2016).
 
\bibitem{100}
  R.~Shaikh,
  Phys.\ Rev.\ D {\bf 100} (2019) no.2,  024028.

	\bibitem{1111} 
  C.~Bambi, K.~Freese, S.~Vagnozzi and L.~Visinelli,
    Phys.\ Rev.\ D {\bf 100}, no. 4, 044057 (2019).	
    
\bibitem{222}
  C.~Y.~Chen,
  arXiv:2004.01440 [gr-qc].
 
  
\bibitem{444}
  L.~Amarilla and E.~F.~Eiroa,
  Phys.\ Rev.\ D {\bf 85} (2012) 064019
  



\bibitem{Narayan} R. Narayan, M. D. Johnson and C. F. Gammie,  Astrophys. J. 885, no. 2, L33 (2019)


\bibitem{Zeng:2020dco}
X.~X.~Zeng, H.~Q.~Zhang and H.~Zhang,
Eur. Phys. J. C \textbf{80}, no.9, 872 (2020)

\bibitem{Falcke:1999pj}
H.~Falcke, F.~Melia and E.~Agol,
Astrophys. J. Lett. \textbf{528} (2000), L13

\bibitem{Bambi:2013nla}
C.~Bambi,
Phys. Rev. D \textbf{87} (2013), 107501

\bibitem{Dokuchaev:2020wqk}
V.~I.~Dokuchaev and N.~O.~Nazarova,
[arXiv:2007.14121 [astro-ph.HE]].

\bibitem{ns} R. Shaikh, P. Kocherlakota, R. Narayan, P.S. Joshi, MNRAS 482, 52 (2019).

\bibitem{Gyulchev:2019tvk}
  G.~Gyulchev, P.~Nedkova, T.~Vetsov and S.~Yazadjiev,
  arXiv:1905.05273 [gr-qc].
  
\bibitem{Shaikh:2019hbm}
R.~Shaikh and P.~S.~Joshi,
JCAP \textbf{10} (2019), 064

\bibitem{Xu:2018wow}
  Z.~Xu, X.~Hou, X.~Gong and J.~Wang,
  JCAP {\bf 1809} (2018) no.09,  038

\bibitem{Allahyari:2019jqz}
A.~Allahyari, M.~Khodadi, S.~Vagnozzi and D.~F.~Mota,
JCAP \textbf{02} (2020), 003

\bibitem{Vincent:2020dij}
F.~H.~Vincent, M.~Wielgus, M.~A.~Abramowicz, E.~Gourgoulhon, J.~P.~Lasota, T.~Paumard and G.~Perrin,
[arXiv:2002.09226 [gr-qc]].

\bibitem{Frost:2020zcy}
T.~C.~Frost and V.~Perlick,
[arXiv:2010.11908 [gr-qc]].

\bibitem{Perlick:2018iye}
V.~Perlick, O.~Y.~Tsupko and G.~S.~Bisnovatyi-Kogan,
Phys. Rev. D \textbf{97}, no.10, 104062 (2018)



\bibitem{Khodadi:2020jij}
M.~Khodadi, A.~Allahyari, S.~Vagnozzi and D.~F.~Mota,
[arXiv:2005.05992 [gr-qc]].

		\bibitem{Bambi:2019tjh}
		C.~Bambi, K.~Freese, S.~Vagnozzi and L.~Visinelli,
	Phys.\ Rev.\ D {\bf 100}, no. 4, 044057 (2019)
		[arXiv:1904.12983 [gr-qc]].
		
		\bibitem{Vagnozzi:2019apd}
		S.~Vagnozzi and L.~Visinelli,
	Phys.\ Rev.\ D {\bf 100}, no. 2, 024020 (2019)
		
		\bibitem{G1}
  G.~Gyulchev, P.~Nedkova, V.~Tinchev and S.~Yazadjiev,
  Eur.\ Phys.\ J.\ C {\bf 78},  544 (2018). 
  
\bibitem{G2}
  G.~Gyulchev, P.~Nedkova, V.~Tinchev and Stoytcho Yazadjiev,
  AIP Conf.\ Proc.\  {\bf 2075},  040005 (2019).

  \bibitem{111} 
  K.~Jusufi, M.~Jamil, P.~Salucci, T.~Zhu and S.~Haroon,
  Phys.\ Rev.\ D {\bf 100}, no. 4, 044012 (2019). 
  
  \bibitem{33}
  S.~Haroon, M.~Jamil, K.~Jusufi, K.~Lin and R.~B.~Mann,
  Phys.\ Rev.\ D {\bf 99} (2019) no.4,  044015.
  
    
\bibitem{shao} S.-W. Wei and  Y.-X. Liu, J. Cosmology Astropart. Phys. {\bf 11}, 063 (2013).
\bibitem{Xu} Z.Xu, X. Hou and J. Wang, J. Cosmology and Astropart. Phy. {\bf 10}, 046 (2018).
\bibitem{Hou:2018avu}
X.~Hou, Z.~Xu and J.~Wang,
JCAP \textbf{12} (2018), 040
\bibitem{Zhou} 
  X.~Hou, Z.~Xu, M.~Zhou and J.~Wang,
  JCAP {\bf 1807}, 015  (2018). 
  
\bibitem{44}
S.~Haroon, K.~Jusufi and M.~Jamil, Universe {\bf 6},  23 (2020).

\bibitem{f1}
M.~Jaroszynski and A.~Kurpiewski,
Astron. Astrophys. \textbf{326}, 419 (1997)

\bibitem{f2}
S.~Nampalliwar, A.~G.~Suvorov and K.~D.~Kokkotas,
Phys. Rev. D \textbf{102}, no.10, 104035 (2020)

\bibitem{f3}
X.~X.~Zeng and H.~Q.~Zhang,
Eur. Phys. J. C \textbf{80}, no.11, 1058 (2020)

\bibitem{f4}
S.~E.~Gralla, D.~E.~Holz and R.~M.~Wald,
Phys. Rev. D \textbf{100}, no.2, 024018 (2019)

\bibitem{Epy1}
https://einsteinpy.org/

\bibitem{Bapat:2020xfa}
S.~Bapat, R.~Saha, B.~Bhatt, S.~Jain, A.~Jain, S.~O.~Vela, P.~Khandelwal, J.~Shivottam, J.~Ma and G.~S.~Ng, \textit{et al.}
[arXiv:2005.11288 [gr-qc]].

\bibitem{Kala:2020prt}
S.~Kala, Saurabh, H.~Nandan and P.~Sharma,
Int. J. Mod. Phys. A \textbf{35} (2020) no.28, 2050177

\bibitem{Konoplya:2019sns}
R.~A.~Konoplya,
Phys. Lett. B \textbf{795} (2019), 1-6
\bibitem{333}
  R.~C.~Pantig and E.~T.~Rodulfo, Chin. J. Phys. (2020); R.~C.~Pantig and E.~T.~Rodulfo,
Chin. J. Phys. \textbf{66} (2020), 691-702;  R.~C.~Pantig et al., [arXiv:2104.04304 [gr-qc]].
\bibitem{Jusufi:2019ltj}
  K.~Jusufi,
 Phys. Rev. D  {\bf 101}, 084055 (2020)
  
\bibitem{Cuadros-Melgar:2020kqn}
B.~Cuadros-Melgar, R.~D.~B.~Fontana and J.~de Oliveira,
Phys. Lett. B \textbf{811}, 135966 (2020)


\bibitem{Li}M. H. Li and  K. C. Yang, Phys. Rev. D 86, 123015 (2012).

\bibitem{K1} V.V. Kiselev (2003) arXiv:gr-qc/0303031.

\bibitem{Xu:2017bpz}
Z.~Xu, J.~Wang and X.~Hou,
Class. Quant. Grav. \textbf{35} (2018) no.11, 115003
\bibitem{Perlick:2015vta} Perlick V., Tsupko O. Y., Bisnovatyi-Kogan G. S., 2015, Phys. Rev. D, 92,
104031
\end{thebibliography}
\end{document}